\documentclass[12pt]{article}
\usepackage[letterpaper, margin = 0.75in]{geometry}
\usepackage{amsmath, amsthm, amssymb, bbm}
\usepackage{graphicx, subcaption, caption, float, adjustbox, psfrag, epsf}
\usepackage{setspace, enumerate, hyperref, xcolor, arydshln, rotating}
\usepackage[ruled,vlined]{algorithm2e}
\usepackage{url}
\usepackage[
  backend=bibtex,
  style=authoryear,
  citestyle=authoryear,
  maxcitenames=2,
  maxbibnames=99
]{biblatex}
\theoremstyle{definition}
\newtheorem{definition}{Definition}

\title{\bf Constrained Minimum Energy Designs}
\author{\\
  Chaofan Huang \hspace{1cm} V. Roshan Joseph \\
  H. Milton Stewart School of Industrial and Systems Engineering, \\
  Georgia Institute of Technology, Atlanta, GA, 30332 \\
  \\
  and \\
  \\
  Douglas M. Ray \\
  US Army-CCDC Armaments Center, \\
  Picatinny Arsenal,  NJ, 07806
}

\begin{document}

\maketitle

\bigskip
\begin{abstract}
Space-filling designs are important in computer experiments, which are critical for building a cheap surrogate model that adequately approximates an expensive computer code. Many design construction techniques in the existing literature are only applicable for rectangular bounded space, but in real world applications, the input space can often be non-rectangular because of constraints on the input variables. One solution to generate designs in a constrained space is to first generate uniformly distributed samples in the feasible region, and then use them as the candidate set to construct the designs. Sequentially Constrained Monte Carlo (SCMC) is the state-of-the-art technique for candidate generation, but it still requires large number of constraint evaluations, which is problematic especially when the constraints are expensive to evaluate. Thus, to reduce constraint evaluations and improve efficiency, we propose the Constrained Minimum Energy Design (CoMinED) that utilizes recent advances in deterministic sampling methods. Extensive simulation results on 15 benchmark problems with dimensions ranging from 2 to 13 are provided for demonstrating the improved performance of CoMinED over the existing methods.


\end{abstract}

\noindent%

\newpage

\section{Introduction}
\label{sec:introduction}

In deterministic computer experiments, we use computer codes to study the input/output relationship of some complex physical, economical, or engineering models, e.g. large eddy simulations for rocket engine injector design \parencite{mak2018surrogate}. However, computer simulations are often time-consuming. Thus, the first step is to build a computationally cheap surrogate model that approximates the expensive computer code using some offline simulation runs \parencite{santner2018cdoe}. 

Space-filling designs are commonly used for constructing the experimental designs where we run the computer simulations. Since we do not have a priori information about the input/output relationship, it is important to have the design points well spread out across the entire design region $\mathcal{X} \subseteq\mathbb{R}^p$. Minimax and maximin are the two popular space-filling measures proposed by \textcite{johnson1990design}. A minimax design aims to minimize the maximum distance from any point in $\mathcal{X}$ to the closest design point, whereas a maximin design maximizes the minimum distance between any two design points. Due to computational tractability, the maximin measure is more commonly used in the literature, which is followed in this paper as well. However, maximin designs are often collapsing, that is, some design points share the same value in one-dimensional projections. Latin hypercube designs \parencite[LHDs;][]{mckay1979lhd} are developed for having good projection of each factor, which can be further improved by integrating it with other space-filling criteria such as maximin \parencite{morris1995MmLHD}. However, maximin LHDs can only ensure good one-dimensional projection and full-dimensional space-fillingness. The maximum projection (MaxPro) designs \parencite{joseph2015maxpro}, on the other hand, are able to achieve good space-filling properties on projections to all subsets of factors. 

\begin{figure}[t!]
    \centering
    \includegraphics[width=0.4\textwidth]{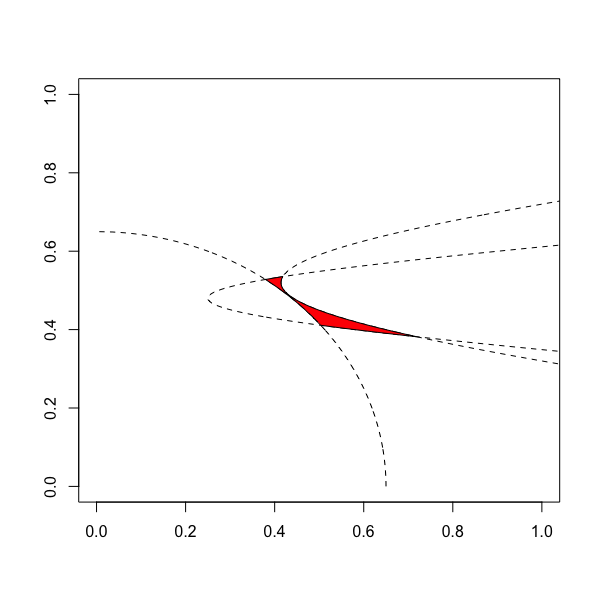}
    \caption{A two-dimensional non-convex, non-rectangular bounded design space $\mathcal{X}$ (in red) with feasibility ratio of only 0.53\% (Table \ref{tab:benchmark_info}) due to the three nonlinear inequality constraints \eqref{eq:MOT}.}
    \label{fig:mot-space}
\end{figure}

Most literature on space-filling designs focus on  bounded rectangular region $\mathcal{X} = \prod_{d=1}^{p}[a_d,b_d]\subseteq\mathbb{R}^{p}$. However, in real world applications such as the welded beam design problem \parencite{dong2018cgo} and the NASA speed reducer design problem \parencite{liu2017cgo}, we frequently need to deal with non-rectangular bounded design space:
\begin{equation}
  \label{eq:space}
  \mathcal{X} = \bigg\{x \in \prod_{d=1}^{p}[a_d,b_d]: g_k(x) \leq 0\; \forall k = 1,\ldots,K \bigg\} \; ,
\end{equation}
where the rectangular shape is jeopardized by the $K$ inequality constraints $\{g_k(x)\leq 0\}_{k=1}^{K}$. For simplicity, let us consider the bounded space of a unit hypercube, that is $a_d = 0,\; b_d = 1\; \forall d = 1,\ldots,p$. This is possible since we can always re-scale the factors. Figure~\ref{fig:mot-space} shows a two-dimensional design space $\mathcal{X}$ obtained by three nonlinear inequality constraints \eqref{eq:MOT}. The non-convex, non-rectangular shape with extremely small feasibility ratio makes it challenging to construct space-filling designs. \par

Two primary approaches are proposed in the literature for constructing designs in non-rectangular design space. One approach is to directly employ general purpose constrained optimization techniques  \parencite{trosset1999cMm, stinstra2003cMm, kang2019}. However, this approach can be computationally very expensive and can be limited by the type of constraints and design properties (such as projections) it can handle. The alternative approach instead relies on a two-step process:
\begin{itemize}
  \item \textbf{Candidate Generation:} generate a large set of uniformly distributed candidates in $\mathcal{X}$.
  \item \textbf{Design Construction:} choose points from the set of candidates by a desired criterion.
\end{itemize}
The flexibility of choosing any design criterion aforementioned in the construction step easily allows for both space-filling and noncollapsing properties in the resulting designs, but how to efficiently generate good quality candidate points remains the \textit{key difficulty} of this approach.  The main objective of this paper is to propose an efficient method to generate good quality candidate points that are suitable for constructing maximin designs.

Several candidates generation methods have been discussed in the literature. If the desired space $\mathcal{X}$ is regularly-shaped, e.g. simplex and circle, where the closed-form inverse Rosenblatt transform exists, we can obtain uniform samples in $\mathcal{X}$ by applying the inverse transform on a set of low-discrepancy sequence in $[0,1]^{p}$ \parencite{Fang1994}. However, we generally cannot compute the inverse Rosenblatt transform for arbitrary irregularly-shaped space $\mathcal{X}$. An alternative solution is to perform acceptance/rejection sampling on a large set of uniformly distributed points in $[0,1]^{p}$, such as grid points \parencite{pratola2017cMm}, Latin hypercube samples \parencite{wu2019cMm}, and quasi-random points \parencite{joseph2016maxpro}. For the design space $\mathcal{X}$ considered in Figure~\ref{fig:mot-space}, given its small feasibility ratio of 0.53\%, on average only 5 out of 1{,}000 samples in the unit hypercube would land in the design space, indicating that the one-step acceptance/rejection approach can be highly inefficient. One remedy is to iterate between acceptance/rejection sampling and candidate augmentation \parencite{draguljic2012cMm}. To benefit from simulated annealing \parencite{kirkpatrick1983sa}, the multi-step acceptance/rejection can be performed on a sequence of shrinking regions \parencite[Subset Simulation;][]{bect2017ss}, and this idea is further improved using the probabilistic constraint, leading to the Sequentially Constrained Monte Carlo \parencite[SCMC;][]{golchi2015scmc,golchi2016scmc}. However, SCMC suffers the same issue of Monte Carlo sampling: many samples are repeated or are very close to each other, which add minimal value for the ultimate goal of constructing a maximin design. Moreover, by having the samples well spread out, fewer proposed samples are required to cover the entire design space, and thus fewer evaluations of the constraints, which is beneficial when the constraints are expensive to evaluate. Minimum Energy Design (MinED) is a state-of-the-art deterministic sampling method for simulating well-spaced samples for any distribution \parencite{joseph2015mined,joseph2019mined}. When the target distribution is uniform, the MinED is equivalent to the maximin design, showing its strong connection to the problem considered in this paper. Thus, by incorporating the probabilistic constraints from SCMC in MinED, we propose the Constrained Minimum Energy Design (CoMinED) as a more efficient approach for generating good quality design candidate samples in arbitrarily constrained space. \par

The paper is organized as follows. Section~\ref{sec:existing_algorithms} reviews the existing candidates generation and designs construction algorithms. Section~\ref{sec:constrained_minimum_energy_design} discusses minimum energy design and proposes the constrained minimum energy design (CoMinED). Section~\ref{sec:simulation_results} demonstrates the improvement of the proposed CoMinED with extensive simulation studies. We conclude the article with some remarks and future research directions in Section~\ref{sec:conclusion}.

\section{Existing Algorithms}
\label{sec:existing_algorithms}

\subsection{Candidate Generation}
\label{subsec:candidates_generation}

\subsubsection{Acceptance/Rejection Sampling}
\label{subsubsec:acceptance_rejection_sampling}

\begin{figure}[t!]
  \centering
    \begin{subfigure}{0.33\textwidth}
      \centering
      \includegraphics[width=0.9\textwidth]{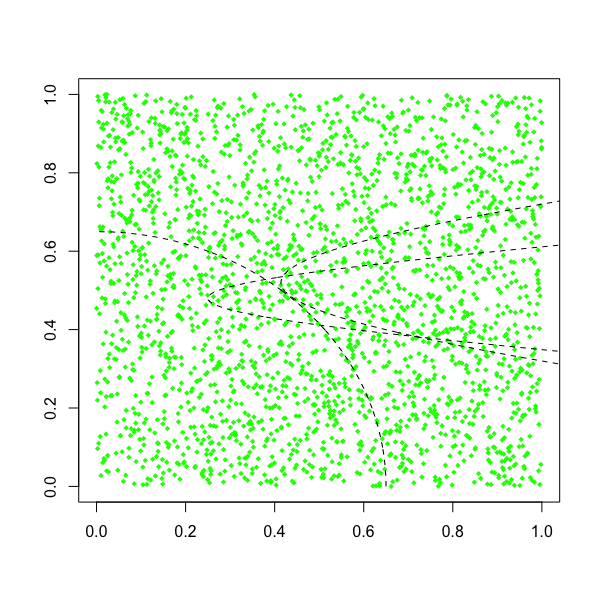}
      \caption{2{,}385 candidate samples}
    \end{subfigure}%
    \begin{subfigure}{0.33\textwidth}
      \centering
      \includegraphics[width=0.9\textwidth]{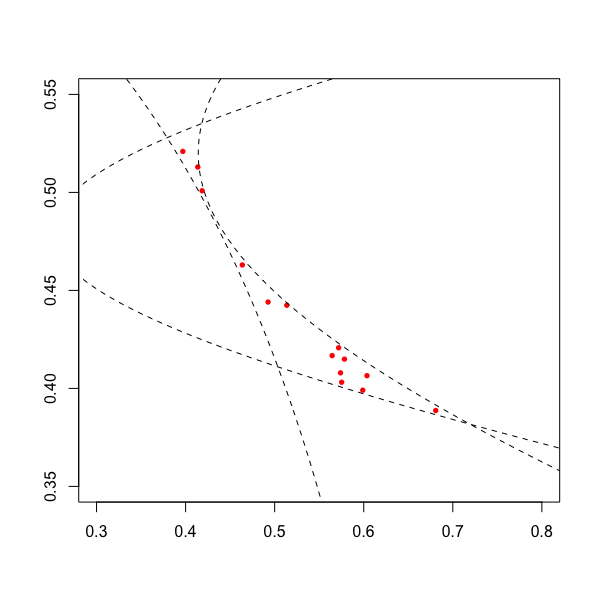}
      \caption{14 feasible samples}
    \end{subfigure}%
  \caption{Left panel shows 2{,}385 randomized Latin hypercube samples from $[0,1]^2$. Right panel shows the 14 feasible candidate samples in $\mathcal{X}$ after applying acceptance/rejection sampling.}
  \label{fig:mot-lhd}
\end{figure}

The simplest approach to generate candidates in any constrained space $\mathcal{X}$ is to first simulate large set of uniformly distributed samples in a rectangular region that contains $\mathcal{X}$, and then apply acceptance/rejection sampling based on the constraints to keep only the feasible samples. Figure~\ref{fig:mot-lhd} shows that only 14 out of the 2{,}385 randomized Latin hypercube samples in $[0,1]^2$ are feasible for the motivation problem in Figure~\ref{fig:mot-space}. The Latin hypercube samples are generated using R package \texttt{lhs} \parencite{carnell2019lhs}. Apart from the low percentage of feasible samples, we can see that these 14 points do not cover the feasible space uniformly well, showing that the one-step acceptance/rejection sampling performs poorly on constrained design problems with very small feasibility ratio. In fact, this issue becomes more serious in higher dimensional problems. 

The inefficiency of the one-step acceptance/rejection sampling results from wasting majority of resources exploring the unit hypercube rather than the target space $\mathcal{X}$. Thus, a $T$-step acceptance/rejection sampling on a sequence of shrinking regions $[0,1]^{p} = \mathcal{X}_0 \supset \mathcal{X}_1 \supset \cdots \supset \mathcal{X}_T = \mathcal{X}$ would allow for exploitation of the important region that is likely feasible. This is known as the subset simulation for estimating the probability of failure in reliability analysis \parencite{bect2017ss}. Consider a non-rectangular bounded design space that is defined by one inequality constraint, $\mathcal{X} = \{x\in[0,1]^{p}:g(x) \leq 0\}$. The subset simulation defines $\{\mathcal{X}_{t}\}_{t=0}^{T}$ by introducing a decreasing sequence of thresholds, $\infty = u_0 > u_1 > \cdots > u_T = 0$, such that $\mathcal{X}_{t} = \{x\in[0,1]^{p}:g(x) \leq u_t\}$. The acceptance/rejection sampling is equivalent to sampling from a indicator function, so we can also view the subset simulation as sampling from the sequence of distributions $\{\mathbbm{1}_{\mathcal{X}_{t}}(x) = \mathbbm{1}(g(x) \leq u_t)\}_{t=0}^{T}$, and Sequential Monte Carlo (SMC) sampling can be applied. The sampling step in SMC is usually by Markov Chain Monte Carlo \parencite[MCMC;][]{robert2013mc}, but poor performances of MCMC on indicator function are observed in practice. One solution is to replace the hard constraint $g(x)\leq 0$ with a probabilistic constraint, leading to the Sequentially Constrained Monte Carlo, which is discussed next.

\subsubsection{Sequentially Constrained Monte Carlo}
\label{subsubsec:sequentially_constrained_monte_carlo}

\begin{algorithm}[t!]
\SetAlgoLined
  \textbf{Design Space:} $\mathcal{X} = \{x\in[0,1]^{p}:g_k(x)\leq 0 \; \forall k = 1,\ldots,K\}$. \\
  \textbf{Initialization:}
  \begin{itemize}
    \setlength{\itemsep}{0pt}
    \setlength{\parskip}{0pt}
    \setlength{\parsep}{0pt}
    \item set the increasing sequence of rigidity parameters $0 = \tau_{0} < \tau_{1} < \cdots < \tau_{T}$.
    \item simulate the initial $M$ samples $\{x_m^{(0)}\}_{m=1}^{M}$ from $[0,1]^{p}$.
  \end{itemize}
  \vspace{-\topsep}
  \vspace{2mm}
  \For{$t = 1,\ldots,T$}{
    $\bullet$ \textbf{Weighting:} compute the importance weight,
    \begin{equation}
      \label{eq:scmc1}
      w_{m}^{(t)} = \rho_{\tau_{t}}(x_{m}^{(t-1)}) / \rho_{\tau_{t-1}}(x_{m}^{(t-1)}) \; ,
    \end{equation}
    where $\rho_{\tau}(\cdot)$ is defined in \eqref{eq:pc3}. Normalize the weight by $\bar{w}_{m}^{(t)} = w_{m}^{(t)} / \sum_{i=1}^{M}w_{i}^{(t)}$. \\
    $\bullet$ \textbf{Resample:} draw $M$ samples $\{y_m^{(t)}\}_{m=1}^{M}$ from $\sum_{m=1}^{M}\bar{w}_{m}^{(t)}\delta(x - x_m^{(t-1)})$.\\
    $\bullet$ \textbf{Sampling:} for $m = 1,\ldots,N$, draw $x_m^{(t)} \sim K_{\sigma^{(t)}}(y_m^{(t)}, \cdot)$ where $K_{\sigma^{(t)}}(y_m^{(t)}, \cdot)$ is a Markov kernel with target distribution $\rho_{\tau_{t}}$ and adaptive scale $\sigma^{(t)}$ where
    \begin{equation}
      \label{eq:scmc2}
      \sigma^{(t)} = \bigg(\mbox{75\% quantile of } \{\min_{j\neq m}\lVert x_{m}^{(t-1)} - x_{j}^{(t-1)}\rVert\}_{m=1}^{M}\bigg)/\sqrt{p} \;. 
    \end{equation} \\
    
  }
  \vspace{2mm}
  \textbf{Return:} all particles $\{x_{m}^{(t)}\}_{m=1}^{M}{}_{t=0}^{T}$ that are in $\mathcal{X}$. \\
 \caption{Adaptive Sequentially Constrained Monte Carlo.}
 \label{algo:scmc}
\end{algorithm}

For any inequality constraint $g(x) \leq 0$, \textcite{golchi2015scmc} proposed the following probabilistic relaxation using the probit function,
\begin{equation}
  \label{eq:pc1}
  \rho_{\tau}(x) = \Phi(-\tau g(x))\; ,
\end{equation}
where $\Phi$ is the standard normal cumulative distribution function and $\tau$ is the parameter that controls the \textit{rigidity} of the constraint. We can see that the function $\rho_{\tau}$ assigns value close to 1 for $x$ that meets the constraint and value close to 0 otherwise. Moreover, in the limit,
\begin{equation}
  \label{eq:pc2}
  \lim_{\tau\to\infty}\rho_{\tau}(x) = \lim_{\tau\to\infty}\Phi(-\tau g(x)) = \mathbbm{1}(g(x) \leq 0)\; .
\end{equation}
The above can be generalized to multiple inequality constraints $\{g_{k}(x)\leq 0\}_{k=1}^{K}$ by 
\begin{equation}
  \label{eq:pc3}
  \rho_{\tau}(x) = \prod_{k=1}^{K}\Phi(-\tau g_{k}(x))\; .
\end{equation}
This leads to the Sequentially Constrained Monte Carlo (SCMC) that replaces the sequence of indicator functions $\{\mathbbm{1}_{\mathcal{X}_{t}}\}_{t=0}^{T}$ in the subset simulation by the sequence of probabilistic constraint functions $\{\rho_{\tau_{t}}\}_{t=0}^{T}$ defined in \eqref{eq:pc3} with an increasing sequence $0 = \tau_{0} < \tau_{1} < \cdots < \tau_{T}$ where $\tau_{T}$ is a large constant, e.g. $10^{6}$. However, in the SCMC algorithm of \textcite{golchi2015scmc}, a pre-fixed normal distribution proposal is used in the Markov kernel of the MCMC step, but how to pick the scale of the normal proposal is difficult for high dimensional problem with small feasible region. Thus, for a more robust comparison to our proposed approach, we improve the SCMC method by allowing adaptation of the Markov kernel. At each iteration, the scale (standard deviation) of the normal proposal is adapted to be the 75\% quantile of the prior step samples' distances to their closest neighbors divided by the square root of the problem dimension. This adaptive kernel shows robust performance for majority of the benchmark problems considered in this paper. Algorithm~\ref{algo:scmc} details the \textit{adaptive SCMC} algorithm for generating large number of uniformly distributed samples from any design space $\mathcal{X}$ with arbitrary number of constraints. 

\begin{figure}[t!]
  \centering
    \begin{subfigure}{0.33\textwidth}
      \centering
      \includegraphics[width=0.9\textwidth]{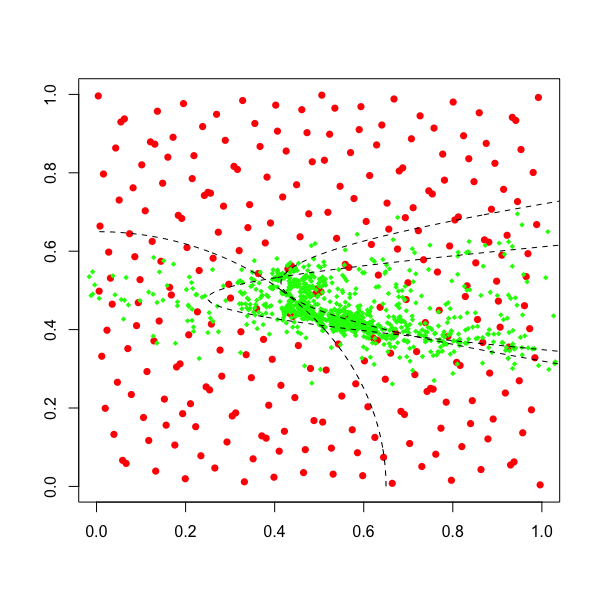}
      \caption{2{,}385 candidate samples}
    \end{subfigure}%
    \begin{subfigure}{0.33\textwidth}
      \centering
      \includegraphics[width=0.9\textwidth]{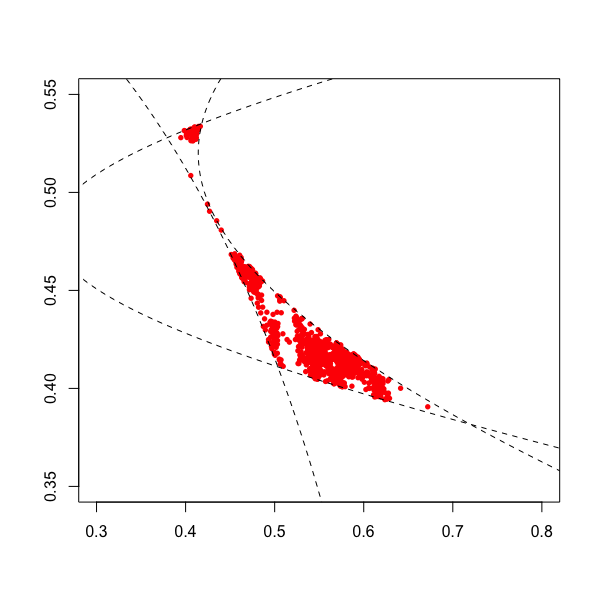}
      \caption{1{,}205 feasible samples}
    \end{subfigure}%
  \caption{Left panel shows 2{,}385 candidate samples from applying adaptive SCMC (Algorithm~\ref{algo:scmc}) on the design space $\mathcal{X}$ defined by \eqref{eq:MOT} with $M = 265$, $T = 8$, and $\{\tau_{t}\}_{t=0}^{8} = [0,e^{1},e^{2},e^{3},e^{4},e^{5},e^{6},e^{7},10^{6}]$. Red circles indicate the initial candidate set of 265 Sobol' points. Right panel shows the 1{,}205 feasible candidate samples.}
  \label{fig:mot-scmc}
\end{figure}

Figure~\ref{fig:mot-scmc} shows the performance of the adaptive SCMC on the motivation problem in Figure~\ref{fig:mot-space} with $M = 265$, $T = 8$, and $\{\tau_{t}\}_{t=0}^{8} = [0,e^{1},e^{2},e^{3},e^{4},e^{5},e^{6},e^{7},10^{6}]$. The 265 Sobol' points in $[0,1]^2$ simulated by the R package \texttt{randtoolbox} \parencite{christophe2019randtoolbox} are used as the initial candidate set. We can see that now with 2{,}385 samples, the adaptive SCMC yields 1{,}205 feasible samples, and they cover the feasible space much better than the the one-step acceptance/rejection sampling on the Latin hypercube samples (Figure~\ref{fig:mot-lhd}), but we can still see large gaps left unexplored in the feasible space.

\subsection{Design Construction}
\label{subsec:designs_construction}
The ultimate goal is to construct an $n$-point design $\mathcal{D}_{n} = \{x_{i}\in\mathcal{X}\}_{i=1}^{n}$ in $\mathcal{X}$ that achieves some good design properties. From the candidate generation step, we obtain a finite set of $N$ ($N \geq n$) candidate points $\mathcal{C}_{N} = \{y_j\in\mathcal{X}\}_{j=1}^{N}$ that are approximately uniformly distributed in $\mathcal{X}$. The next step is to find the $n$ samples from the candidate set that maximize a desired design criterion $\psi$, that is to solve 
\begin{equation}
  \label{eq:dc1}
  \arg\max_{\mathcal{D}_{n}\subseteq \mathcal{C}_{N}} \psi(\mathcal{D}_{n}) \; .
\end{equation}
In the case of the maximin design, $\psi(\mathcal{D}_{n}) = \min_{x_i,x_j\in \mathcal{D}_{n}; i\neq j}\lVert x_i - x_j\rVert_{2}$ where $\lVert \cdot \rVert_{2}$ is the Euclidean distance. Many stochastic optimization algorithms have been developed to efficiently address the combinatorial optimization problem in \eqref{eq:dc1}, including local search, threshold accepting, simulated annealing \parencite{morris1995MmLHD}, enhanced stochastic evolutionary \parencite{jin2003MmLHD,wu2019cMm}. See \textcite{fang2005design} Chapter 4 for a detailed review of the above methods. On the other hand, \textcite{kennard1969cdoe} proposed a one-point-at-a-time greedy algorithm for solving \eqref{eq:dc1}. The idea is that by having a $m$-point design $\mathcal{D}_{m}$ ($m < n$), we generate the $(m+1)$-th point by
\begin{equation}
  \label{eq:dc2}
  x_{m+1} = \arg\max_{x\in \mathcal{C}_{N}\backslash \mathcal{D}_{m}}\psi(\mathcal{D}_{m}\cup\{x\}) \; .
\end{equation}
The one-point-at-a-time greedy procedure is also employed in the R package \texttt{mined} \parencite{wang2019mined} for generating minimum energy design and the R package \texttt{MaxPro} \parencite{ba2018maxpro} for design augmentation. Although the one-point-at-a-time greedy algorithm results in a local optimum, it is efficient and shows good empirical performance in practice. We also employ this greedy procedure in the constrained minimum energy design discussed in the next section.

\section{Constrained Minimum Energy Design}
\label{sec:constrained_minimum_energy_design}

\subsection{Minimum Energy Design}
\label{subsec:minimum_energy_design}

We begin by formally defining the minimum energy design (MinED).

\begin{definition}{\textbf{\parencite[Minimum Energy Design;][]{joseph2015mined}}}
\label{df:minimum_energy_designs}
Let $\pi$ be the target probability density function. An n-point minimum energy design of $\pi$ is the optimal solution of 
\begin{equation}
  \label{eq:mined1}
  \arg\min_{\mathcal{D}_n\in\mathbb{D}_n}\sum_{\substack{x_i,x_j\in\mathcal{D}_n\\i\neq j}}\frac{q(x_i)q(x_j)}{\lVert x_i - x_j\rVert_{2}}\; ,
\end{equation}
where $\mathbb{D}_{n} = \{\{x_i\}_{i=1}^{n}:x_i\in\mathbb{R}^{p}\}$ is the set of all unordered n-tuple in $\mathbb{R}^{p}$. $q(\cdot) = 1/\pi^{1/(2p)}(\cdot)$ is the charge function and $\lVert \cdot \rVert_{2}$ is the Euclidean distance. Under the proposed charge function, the limiting distribution of the design points converges to $\pi$. 
\end{definition}
\noindent However, the optimization problem in \eqref{eq:mined1} is difficult to solve and numerically unstable. To circumvent this issue, \textcite{joseph2019mined} recognize that \eqref{eq:mined1} is closely related to
\begin{equation}
  \label{eq:mined2}
  \arg\min_{\mathcal{D}_n\in\mathbb{D}_n}\bigg[\sum_{\substack{x_i,x_j\in\mathcal{D}_n\\i\neq j}}\bigg(\frac{q(x_i)q(x_j)}{\lVert x_i - x_j\rVert_{2}}\bigg)^{k}\bigg]^{1/k}
\end{equation}
for $k>0$. As $k\to\infty$, the optimization problem becomes
\begin{equation}
  \label{eq:mined3}
  \arg\min_{\mathcal{D}_n\in\mathbb{D}_n}\max_{\substack{x_i,x_j\in\mathcal{D}_n\\i\neq j}}\frac{q(x_i)q(x_j)}{\lVert x_i - x_j\rVert_{2}}\; .
\end{equation}
By substituting $q(\cdot) = 1/\pi^{1/(2p)}(\cdot)$ into \eqref{eq:mined3}, we have
\begin{equation}
  \begin{aligned}
    \label{eq:mined4}
    & \arg\min_{\mathcal{D}_n\in\mathbb{D}_n}\max_{\substack{x_i,x_j\in\mathcal{D}_n\\i\neq j}}\frac{1}{\pi^{1/(2p)}(x_i)\pi^{1/(2p)}(x_j)\lVert x_i - x_j\rVert_{2}} \\
    = & \arg\max_{\mathcal{D}_n\in\mathbb{D}_n}\min_{\substack{x_i,x_j\in\mathcal{D}_n\\i\neq j}}\pi^{1/(2p)}(x_i)\pi^{1/(2p)}(x_j)\lVert x_i - x_j\rVert_{2} \\
    = & \arg\max_{\mathcal{D}_n\in\mathbb{D}_n}\min_{\substack{x_i,x_j\in\mathcal{D}_n\\i\neq j}}\gamma^{1/(2p)}(x_i)\gamma^{1/(2p)}(x_j)\lVert x_i - x_j\rVert_{2} \\
    = & \arg\max_{\mathcal{D}_n\in\mathbb{D}_n}\min_{\substack{x_i,x_j\in\mathcal{D}_n\\i\neq j}}\frac{1}{2p}\log\gamma(x_i) + \frac{1}{2p}\log\gamma(x_j) + \log \lVert x_i - x_j\rVert_{2} \; , \\
  \end{aligned}
\end{equation}
where $\gamma \propto \pi$ is the unnormalized probability density function. Now we only need to work with the log-unnormalized density and therefore, the objective function is numerically more stable. Intuitively, \eqref{eq:mined4} wants the design points to be as far apart as possible while are still placed in the high density regions. If we take $\pi = \mbox{Uniform}[0,1]^{p}$, then \eqref{eq:mined4} reduces to 
\begin{equation}
  \label{eq:mined5}
  \arg\max_{\mathcal{D}_n\in\mathbb{D}_n^{u}}\min_{\substack{x_i,x_j\in\mathcal{D}_n\\i\neq j}}\lVert x_i - x_j\rVert_{2} \; ,
\end{equation}
where $\mathbb{D}_n^{u} = \{\{x_i\}_{i=1}^{n}:x_i\in[0,1]^{p}\}$ is the set of all unordered n-tuple in $[0,1]^{p}$. \eqref{eq:mined5} is the same optimization problem for the maximin design in the unit hypercube \parencite{johnson1990design}. \textcite{joseph2019mined} further propose a generalized distance, 
\begin{equation}
  \label{eq:mined6}
  \lVert u\rVert_{s} = \bigg(\frac{1}{p}\sum_{l=1}^{p}|u_{l}|^{s}\bigg)^{1/s},\;  s > 0.
\end{equation}
Under the distance measure defined in \eqref{eq:mined6}, the limiting distribution of the MinED points converge to $\pi$ for all $s > 0$. When $s \to 0$, the distance measure converge to $\lVert u\rVert_{0} = \prod_{l=1}^{p}|u_{l}|^{1/p}$. If $\pi$ is the uniform distribution, MinED with $\lVert \cdot\rVert_{0}$ is the limiting case of the MaxPro design \parencite{joseph2019mined}, showing that noncollapsing property can also be easily achieved by carefully choosing the distance measure.

\subsection{Constrained Minimum Energy Design}
\label{subsec:constrained_minimum_energy_design}
Now consider the case that we need to generate MinED for $\gamma \propto \pi$, an unnormalized probability density function, in some non-rectangular bounded space $\mathcal{X}=\{x\in[0,1]^{p}:g_{k}(x)\leq 0\; \forall k = 1,\ldots,K\}$, then the optimization problem becomes
\begin{equation}
  \label{eq:comined1}
  \arg\max_{\mathcal{D}_n\in\mathbb{D}_n^{\mathcal{X}}}\min_{\substack{x_i,x_j\in\mathcal{D}_n\\i\neq j}}\frac{1}{2p}\log\gamma(x_i) + \frac{1}{2p}\log\gamma(x_j) + \log \lVert x_i - x_j\rVert_{s} \; ,
\end{equation}
where $\mathbb{D}_n^{\mathcal{X}} = \{\{x_i\}_{i=1}^{n}:x_i\in\mathcal{X}\}$ is the set of all unordered n-tuple in $\mathcal{X}$. However, constraint optimization is generally hard to solve, especially when some of the constraint functions are nonlinear. Similar to Sequentially Constrained Monte Carlo, we can simplify the optimization problem \eqref{eq:comined1} by introducing the probabilistic relaxation $\rho_{\tau}$, \eqref{eq:pc3}, for the inequality constraints $\{g_{k}\}_{k=1}^{K}$, leading to the constrained minimum energy design (CoMinED) defined below.
\begin{definition}{\textbf{(Constrained Minimum Energy Design)}}
\label{df:constrained_minimum_energy_design}
Let $\gamma \propto \pi$ be the target unnormalized probability density function. An n-point minimum energy design of $\pi$ in any non-rectangular bounded space $\mathcal{X}=\{x\in[0,1]^{p}:g_{k}(x)\leq 0\; \forall k = 1,\ldots,K\}$ is the optimal solution of 
\begin{equation}
  \label{eq:comined2}
  \arg\max_{\mathcal{D}_n\in\mathbb{D}_n^{u}}\min_{\substack{x_i,x_j\in\mathcal{D}_n\\i\neq j}}\frac{1}{2p}\log\tilde{\gamma}_{\tau}(x_i) + \frac{1}{2p}\log\tilde{\gamma}_{\tau}(x_j) + \log \lVert x_i - x_j\rVert_{s} \; ,
\end{equation}
where $\mathbb{D}^{u}_{n} = \{\{x_i\}_{i=1}^{n}:x_i\in [0,1]^{p}\}$ is the set of all unordered n-tuple in unit hypercube, $\lVert \cdot \rVert_{s}$ is the distance measure function defined in \eqref{eq:mined6}, and
\begin{equation}
  \label{eq:comined3}
  \tilde{\gamma}_{\tau}(\cdot) = \gamma(\cdot) \times \rho_{\tau}(\cdot) = \gamma(\cdot) \prod_{k=1}^{K}\Phi(-\tau g_{k}(\cdot)) \; ,
\end{equation}
where $\tau$ controls the rigidity of the constraints. As $\tau\to\infty$, \eqref{eq:comined2} is equivalent to \eqref{eq:comined1} in the limit, and $\tau = 10^{6}$ is sufficient to achieve the limit numerically in practice, provided that the constraints are properly scaled, a point that we will discuss in detail in Section~\ref{sec:simulation_results}.
\end{definition}

Although the CoMinED is applicable for any distribution $\pi$, in this paper, we mainly focus on $\pi = \mbox{Uniform}[0,1]^{p}$ for a direct comparison to the existing methods for generating space-filling design in non-rectangular bounded regions. As pointed out in MinED \parencite{joseph2015mined,joseph2019mined}, solving the optimization directly using nonlinear programming solver is difficult and computationally expensive. The proposed remedy is to (i) generate the design from a set of candidate samples, and (ii) apply simulated annealing on $\tau$ by starting with an ``easier" problem and slowly increasing the rigidity of the constraints, which is also employed in the Sequentially Constrained Monte Carlo. Suppose we do a $T$ step-simulated annealing, we need to first define the increasing sequence of rigidity parameters $0 = \tau_0 < \tau_1 < \cdots < \tau_T = 10^{6}$. At each step, we generate the $n$-point intermediate CoMinED as follows. Let $\tau_{t}$ be the rigidity parameter, $\mathcal{C}^{t} = \{y_{j}^{t}\}_{j=1}^{N_{t}}$ be the candidate samples, and $\mathcal{D}^{t} = \{x_{i}^{t}\}_{i=1}^{n} \subseteq \mathcal{C}^{t}$ be the CoMinED at the $t$-th step. To construct $\mathcal{D}^{t+1}$, we first augment the candidate samples to $\mathcal{C}^{t+1}$ by including the linear combinations of nearby points in $\mathcal{D}^{t}$ which we call adaptive lattice grid refinement, and then apply the one-point-at-a-time greedy algorithm \eqref{eq:dc2} to solve \eqref{eq:comined2} with $\tau_{t+1}$ as the rigidity parameter. Algorithm~\ref{algo:comined} presents the detail procedures of generating CoMinED. 

\begin{algorithm}[t!]
\SetAlgoLined
  \textbf{Design Space:} $\mathcal{X} = \{x\in[0,1]^{p}:g_k(x)\leq 0 \; \forall k = 1,\ldots,K\}$. \\
  \textbf{Initialization:}
  \begin{itemize}
    \setlength{\itemsep}{0pt}
    \setlength{\parskip}{0pt}
    \setlength{\parsep}{0pt}
    \item set the increasing sequence of rigidity parameters $0 = \tau_{0} < \tau_{1} < \cdots < \tau_{T} = 10^{6}$.
    \item set the number of nearest neighbors $Q$ to consider for the candidate augmentation.
    \item simulate $N_{1} > n$ (prime number) lattice points $\{y_j^{(1)}\}_{j=1}^{N_{1}}$ from $[0,1]^{p}$ as the initial set of candidate samples $\mathcal{C}^{1}$. 
  \end{itemize}
  \vspace{-\topsep}
  \vspace{2mm}
  \For{$t = 1,\ldots,T$}{
    $\bullet$ \textbf{Construction:} solve \eqref{eq:comined2} with $\tau = \tau_{t}$ by one-point-at-a-time greedy algorithm \eqref{eq:dc2} to obtain the CoMinED $\mathcal{D}^{t} = \{x_{i}^{t}\}_{i=1}^{n}$, i.e., with $\{x_{1}^{t},\ldots,x_{m}^{t}\}$, $x_{m+1}^{t}$ is given by 
    \begin{equation}
      \label{eq:comined4}
      \begin{aligned}
      x_{m+1}^{t} = \arg\max_{x\in \mathcal{C}^{t}\backslash\{x_{l}^{t}\}_{l=1}^{m}}\min_{i=1:m}\frac{1}{2p}\sum_{k=1}^{K}\log\Phi(-\tau_{t}g_{k}(x)) & + \\
      \frac{1}{2p}\sum_{k=1}^{K}\log\Phi(-\tau_{t}g_{k}(x_i)) &+ \log \lVert x_i - x_j\rVert_{s} \; .
      \end{aligned}
    \end{equation}
    \If{$t < T$}{
      $\bullet$ \textbf{Augmentation:} augment the set of candidate samples $\mathcal{C}^{t+1} = \mathcal{C}^{t} \cup \tilde{\mathcal{C}}^{t}$ where $\tilde{\mathcal{C}}^{t}$ is the set of linear combinations of nearby points in $\mathcal{D}^{t}$. We construct $\tilde{\mathcal{C}}^{t}$ as follows.
      \For{$i = 1,\ldots,n$}{
        $\bullet$ find the $Q$ nearest neighbors of $x_{i}^{t}$ in $\mathcal{D}^{t}$. \\
        $\bullet$ for each nearest neighbor $\tilde{x}_{i,q}$ ($q = 1,\ldots,Q$), compute the mid-point 
        \begin{equation}
          \label{eq:comined5}
            \tilde{y}^{(m)}_{i,q} = x_{i} + \frac{1}{2}(\tilde{x}_{i,q} - x_{i}) = \frac{x_{i}+\tilde{x}_{i,q}}{2} \; ,
        \end{equation}
        and the reflection mid-point 
        \begin{equation}
          \label{eq:comined6}
          \tilde{y}^{(r)}_{i,q} = x_{i} - \frac{1}{2}(\tilde{x}_{i,q} - x_{i}) = \frac{3x_{i} - \tilde{x}_{i,q}}{2}\; .
        \end{equation}
        $\bullet$ Update $\tilde{\mathcal{C}}^{t} = \tilde{\mathcal{C}}^{t} \cup \{\tilde{y}^{(m)}_{i,q},\tilde{y}^{(r)}_{i,q}\}$.
      }
      Remove repeated points in $\tilde{\mathcal{C}}^{t}$, and only keep points in $\tilde{\mathcal{C}}^{t}$ that are not in $\mathcal{C}^{t}$. 
    }
  }
  \vspace{2mm}
  \textbf{Return:} 1. feasible candidate samples $\{y\in \mathcal{C}^{T}:y\in\mathcal{X}\}$ and 2. the CoMinED $\mathcal{D}^{T}$. \\
 \caption{Algorithm for Generating $n$-point CoMinED.}
 \label{algo:comined}
\end{algorithm}

\begin{figure}[t!]
  \centering
  \begin{subfigure}{0.33\textwidth}
      \centering
      \includegraphics[width=0.9\textwidth]{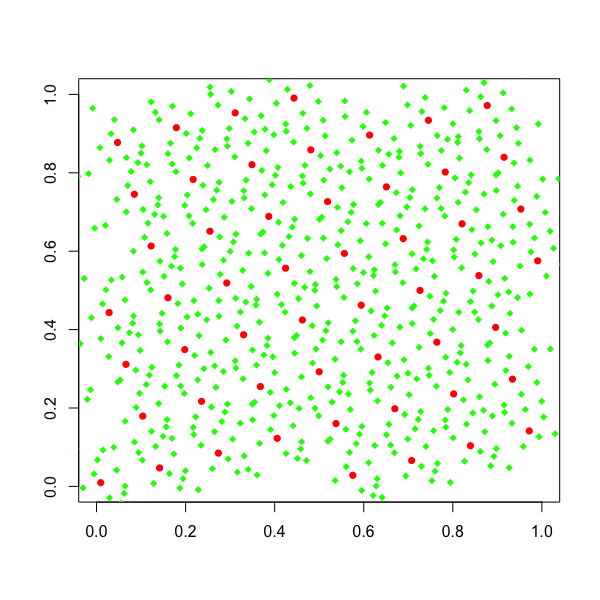}
      \caption{Maximin LHDs}
  \end{subfigure}%
  \begin{subfigure}{0.33\textwidth}
      \centering
      \includegraphics[width=0.9\textwidth]{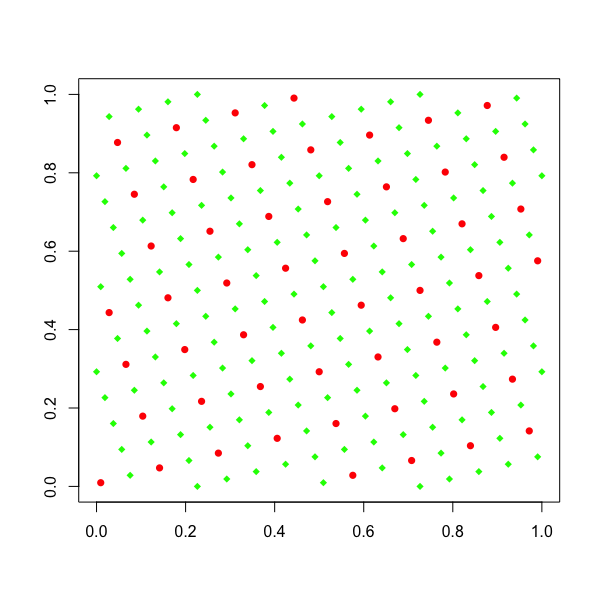}
      \caption{ALGR}
  \end{subfigure}%
  \caption{One step candidate augmentation (in green diamonds) on 53 lattice points (in red circles) by 11 maximin LHDs in local regions (left panel) and adaptive lattice grid refinement (ALGR) considering 11 nearest neighbors (right panel).}
  \label{fig:mmlhd_vs_algr}
\end{figure}

Now let us discuss the advantage of the proposed adaptive lattice grid refinement over the local maximin LHDs for candidate augmentation. In the MinED algorithm, \textcite{joseph2019mined} augment the candidate points by maximin LHDs in the local region of each MinED point, where the local region is defined as the hypercube inscribed in the ball with center being the MinED point and radius being the distance to its nearest neighbor in the design. The left panel of Figure~\ref{fig:mmlhd_vs_algr} shows the candidate augmentation by the local maximin LHDs. The initial set of points (red circles) are generated using ``Lattice" function in R package \texttt{mined} \parencite{wang2019mined} and maximin LHDs are generated using ``maximinLHS" function in R package \texttt{lhs} \parencite{carnell2019lhs}. We see that (i) some of the augmented candidate samples are arbitrarily close because local regions are overlapping, and (ii) some regions are left unexplored since the proposed local hypercubes cannot fill the space fully. Given the aforementioned shortcoming of the local maximin LHDs, we propose the adaptive lattice grid refinement (ALGR) for a better ``space-filling" candidate augmentation. The ALGR is based on the good rank-1 lattice rule \parencite{nuyens2006lattice,nuyens2007lattice}, one popular type of quasi-Monte Carlo (QMC) methods. Different from other QMC samples such as Halton' and Sobol' points \parencite{niederreiter1992qmc}, lattice points possess a grid structure (see the red circles of Figure~\ref{fig:mmlhd_vs_algr}) that make them advantageous for candidate augmentation. Under the grid structure, all lattice points share the same distance $\delta$ to their nearest neighbors. Augmenting the candidate samples by the mid-point \eqref{eq:comined5} and reflection mid-point \eqref{eq:comined6}, we can ensure that the minimal interpoint spacing of the new candidate samples is $\delta/2$. If we do a $T$-step simulated annealing, then the minimum interpoint distance for the final set of candidates would be $\delta/2^{T}$, which agrees with the minimum interpoint distance constraints used in bridge design \parencite{jones2015bridge}. 

The right panel of Figure~\ref{fig:mmlhd_vs_algr} shows the candidate augmentation of the ALGR, which is better space-filling than the local maximin LHDs augmentation. Furthermore, the ALGR exhibits a good trade-off between exploration and exploitation. It starts with a space-filling but sparse grid as the candidate set for good exploration of the whole hypercube. As the rigidity parameter $\tau$ increases, the intermediate CoMinED would only occupy the key regions, leading to refinement of the lattice grid in those regions exclusively for exploitation, and that is why we call it the adaptive lattice grid refinement. 


\begin{figure}[t!]
  \centering
    \begin{subfigure}{0.33\textwidth}
      \centering
      \includegraphics[width=0.9\textwidth]{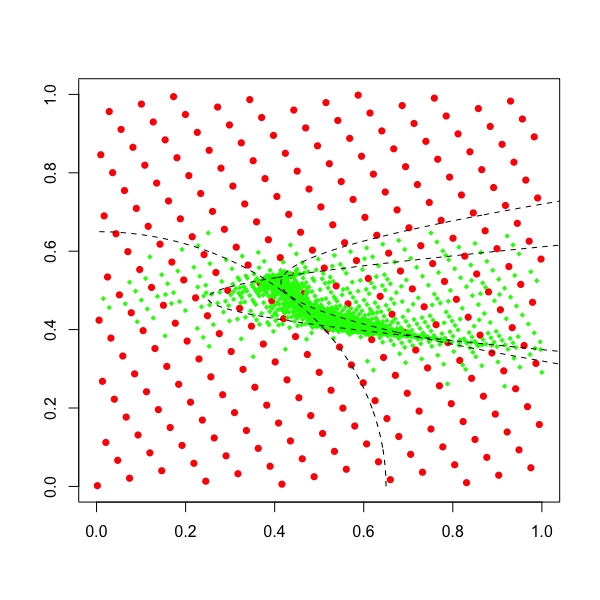}
      \caption{2{,}155 candidate samples}
    \end{subfigure}%
    \begin{subfigure}{0.33\textwidth}
      \centering
      \includegraphics[width=0.9\textwidth]{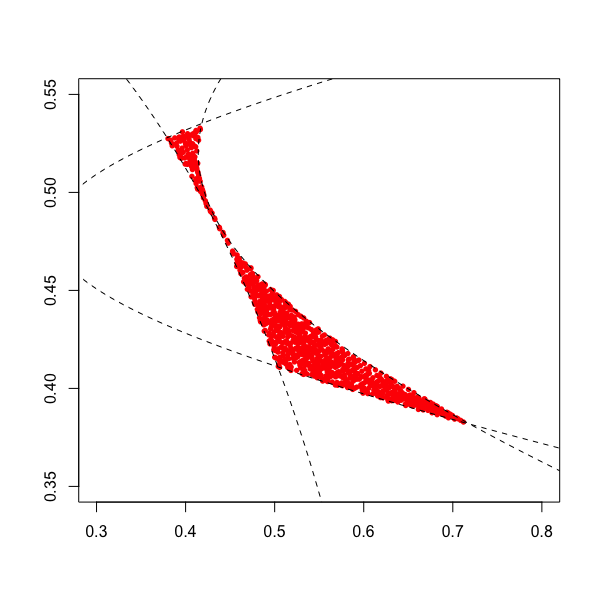}
      \caption{915 feasible samples}
    \end{subfigure}%
  \caption{Left panel shows 2{,}155 candidate samples from applying Algorithm~\ref{algo:comined} to generate $n=53$ points CoMinED on the design space $\mathcal{X}$ defined by \eqref{eq:MOT} with $Q = 5$, $N_{1} = 263$, $T = 8$, and $\{\tau_{t}\}_{t=0}^{8} = [0,e^{1},e^{2},e^{3},e^{4},e^{5},e^{6},e^{7},10^{6}]$. Red circles indicate the initial candidate set of 263 lattice points. Right panel shows the 915 feasible candidate samples.}
  \label{fig:mot-comined}
\end{figure}

Figure~\ref{fig:mot-comined} shows the candidate samples of applying Algorithm~\ref{algo:comined} to generate $n=53$ points CoMinED with $Q=5$ nearest neighbors and $N_{1} = 263$ (the largest prime number that is less than $Qn = 265$) lattice points as the initial candidate set in $T=8$ steps. With 2{,}155 evaluations of the constraints, it yields 915 feasible samples that cover the design space uniformly well: almost no gap spotted visually in the right panel of Figure~\ref{fig:mot-comined}, showing its significant improvement over one-step acceptance/rejection sampling on Latin hypercube samples (Figure~\ref{fig:mot-lhd}) and the adaptive Sequentially Constrained Monte Carlo approach (Figure~\ref{fig:mot-scmc}), which both conduct 2{,}385 evaluations of the constraints to generate the feasible samples. See Figure~\ref{fig:mot-comined_evolution} in Appendix~\ref{appendix:simulation} for the 8-step evolution of the CoMinED.

\section{Simulation Results}
\label{sec:simulation_results}

\begin{table}[t!]
  \centering
  \begin{tabular}{|c|cccc|}
  \hline
  Problem & Dimension ($p$) & No. of LIC & No. of NIC & Feasibility Ratio \\
  \hline
  MOT \eqref{eq:MOT} & 2 & 0 & 3 & 0.0053 \\
  G01 \eqref{eq:G01} & 13 & 9 & 0 & 0.0000 \\
  G04 \eqref{eq:G04} & 5 & 0 & 6 & 0.2696 \\
  G06 \eqref{eq:G06} & 2 & 0 & 2 & 0.0001 \\
  G07 \eqref{eq:G07} & 10 & 3 & 5 & 0.0000 \\
  G08 \eqref{eq:G08} & 2 & 0 & 2 & 0.0086 \\
  G09 \eqref{eq:G09} & 7 & 0 & 4 & 0.0053 \\
  G10 \eqref{eq:G10} & 8 & 3 & 3 & 0.0000 \\
  IBD \eqref{eq:IBD} & 4 & 0 & 3 & 0.0015 \\
  PVD \eqref{eq:PVD} & 4 & 3 & 1 & 0.4032 \\
  SRD \eqref{eq:SRD} & 7 & 0 & 11 & 0.0019 \\
  TSD \eqref{eq:TSD} & 3 & 1 & 3 & 0.0075 \\
  TTD \eqref{eq:TTD} & 2 & 0 & 3 & 0.2179 \\
  WBD \eqref{eq:WBD} & 4 & 1 & 5 & 0.0010 \\
  SCBD \eqref{eq:SCBD} & 10 & 0 & 11 & 0.0005 \\
  \hline
  \end{tabular}
  \caption{Basic information for the constraints of the benchmark problems. LIC stands for linear inequality constraints and NIC stands for nonlinear inequality constraints. Feasibility ratio is estimated using $10^{7}$ Sobol' points in $[0,1]^{p}$ by R package \texttt{randtoolbox} \parencite{christophe2019randtoolbox}.}
  \label{tab:benchmark_info}
\end{table}

In this section, we report the simulation results of applying CoMinED (Algorithm~\ref{algo:comined}) and adaptive SCMC (Algorithm~\ref{algo:scmc}) to 15 benchmark problems with dimensions ranging from 2 to 13 that are popular in the constrained Bayesian optimization literature \parencite{liu2017cgo,dong2018cgo,chaiyotha2020cgo,tao2020cgo}. Since the one-step acceptance/rejection sampling on
Latin hypercube samples is at a clear disadvantage compared to both adaptive SCMC and CoMinED, we do not include the acceptance/rejection sampling simulation results in this section, but are provided in Appendix~\ref{appendix:simulation} for the interested readers. Table~\ref{tab:benchmark_info} provides some basic information of the 15 problems, and their formulations are provided in Appendix~\ref{appendix:benchmark_problems}. All problems are re-scaled to be in the unit hypercube, and the design measures are also compared under the scaling to $[0,1]^{p}$. \par

Throughout the simulations, CoMinED is ran with $s=2$ for the distance measure \eqref{eq:mined6}, the euclidean distance. Also, we take $N_1$, the number of the initial candidate samples, to be the largest prime number that is less than the product of the number of CoMinED points $n$ and the number of neighbors to be considered for candidate augmentation $Q$. For ease of comparison, we fix $T=8$ and the set of rigidity parameters $\{\tau_{t}\}_{t=0}^{8} = [0,e^{1},e^{2},e^{3},e^{4},e^{5},e^{6},e^{7},10^{6}]$, which shows stable performance on both CoMinED and adaptive SCMC. Ways of choosing $Q$ and the rigidity parameters $\tau_{t}$'s are discussed in Appendix~\ref{appendix:comined_details}. Thus, other than the number of design points $n$, $Q$ is the only free parameter that we alter during the simulation for CoMinED. \par

Let us now consider the setting for the adaptive SCMC comparison. Given that we cannot control the total number of constraint evaluations in CoMinED as the repeated samples from augmentation are discarded, we choose the number of samples per iteration in adaptive SCMC to be
\begin{equation}
  \label{eq:scmc_M}
  M = \max\bigg\{nQ, \bigg\lceil\frac{N_{T}}{T+1}\bigg\rceil\bigg\}\;,
\end{equation}
such that (i) $M$ is larger than the number of initial samples of CoMinED ($N_1 < nQ$), and (ii) the total adaptive SCMC samples $M(T+1)$ is larger than the total CoMinED samples $N_{T}$. This puts CoMinED in a slightly \textit{disadvantageous} position in the comparison for assuredly demonstrating its effectiveness over the adaptive SCMC. 

Since CoMinED is a deterministic algorithm by the initial candidate of lattice points and the one-point-at-a-time greedy algorithm for designs construction, only one simulation run is performed. For the adaptive SCMC, 50 runs are used for the comparison. Source codes and tutorials can be found at \url{https://github.com/BillHuang01/CoMinED}.

\subsection{Motivation Example}
\label{subsec:motivation}

Let us first consider the two-dimensional motivation example presented in Figure~\ref{fig:mot-space}. 
\begin{equation}
  \label{eq:MOT}
    \begin{aligned}
      &g_{1}(x) = x_{1} - \sqrt{50(x_{2}-0.52)^2+2} + 1 \leq 0 \\
      &g_{2}(x) = \sqrt{120(x_{2}-0.48)^2+1) - 0.75 - x_{1}} \leq 0 \\
      &g_{3}(x) = 0.65^2 - x_{1}^2 - x_{2}^2 \leq 0 \\
      \mbox{where} \quad & 0 \leq x_{i} \leq 1\; (i=1,2).
    \end{aligned}
\end{equation}
We compare the performance of CoMinED and adaptive SCMC on generating a 53-point design from their feasible candidate set under the following settings: CoMinED is ran with $Q = 5, 11, 17$, and the corresponding adaptive SCMC comparison is ran with $M$ computed by \eqref{eq:scmc_M}. \par

\begin{figure}[t!]
  \centering
  \begin{subfigure}{0.24\textwidth}
    \centering
    \includegraphics[width=0.9\textwidth]{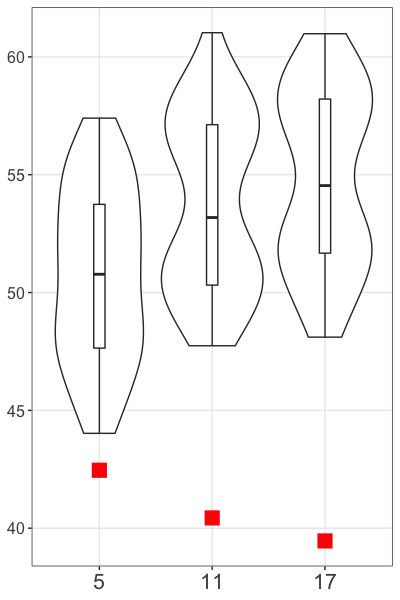}
    \caption{Feasible Ratio}
  \end{subfigure}%
  \begin{subfigure}{0.24\textwidth}
    \centering
    \includegraphics[width=0.9\textwidth]{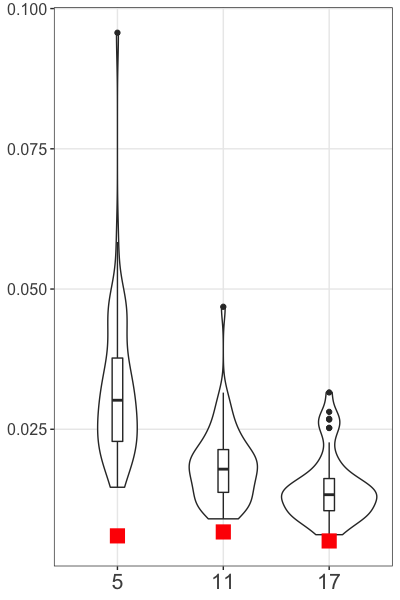}
    \caption{Fill Distance}
  \end{subfigure}%
  \begin{subfigure}{0.24\textwidth}
    \centering
    \includegraphics[width=0.9\textwidth]{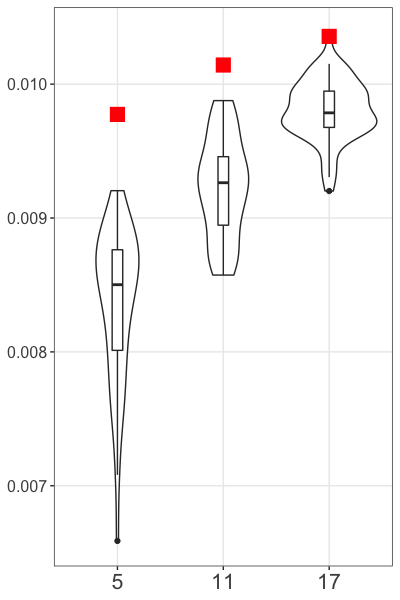}
    \caption{Maximin}
  \end{subfigure}%
  \begin{subfigure}{0.24\textwidth}
    \centering
    \includegraphics[width=0.9\textwidth]{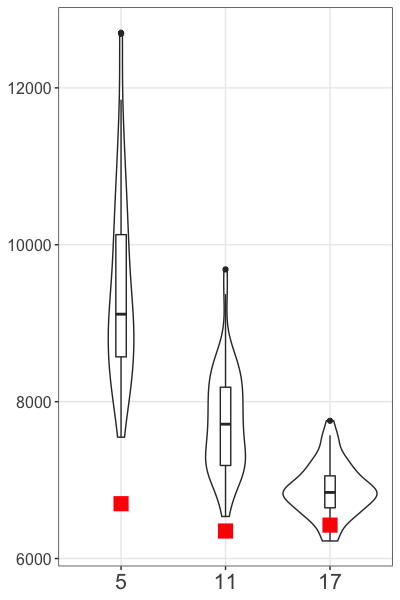}
    \caption{MaxPro}
  \end{subfigure}%
  \caption{Comparisons of the candidates quality and the resulted 53-point design from applying CoMinED (red squares) and adaptive SCMC (violin plots over 50 runs) on the motivation problem \eqref{eq:MOT}. CoMinED is ran with $Q=5,11,17$, and the corresponding adaptive SCMC comparison is ran with $M$ computed by \eqref{eq:scmc_M}. For feasible ratio and maximin measure, the larger the better, and for the fill distance and MaxPro measure, the smaller the better.}
  \label{fig:moto}
\end{figure}

First, let us look at the quality of the feasible samples. Two metrics are considered for the evaluation. One is the \textit{feasible ratio}, the percentage of total samples that are feasible. The larger the feasible ratio, the better the algorithm identifying the feasible region for exploitation. However, the feasible ratio alone could be misleading, as we can always restrict the sampling in an arbitrarily small ball around each feasible point, yielding many samples in the feasible region but also leaving unexplored gaps. For example, Figure~\ref{fig:mot-scmc} shows that 50.5\% ($1{,}205/2{,}385$) of the total adaptive SCMC samples are feasible, but it does not cover the feasible region well as the CoMinED does (Figure~\ref{fig:mot-comined}), which has the feasible ratio of only 42.0\%. Thus, the \textit{fill distance}, the largest distance of any point in $\mathcal{X}$ to the closest feasible samples, is proposed as the other metric to assess how good the algorithm explores the feasible region completely. The smaller the fill distance the better. In simulation, the fill distance is approximated via $10^{4}$ feasible samples from acceptance/rejection sampling on a very large set of Sobol' points in unit hypercube. From Figure~\ref{fig:moto}, we can see that CoMinED exhibits significant improvement in the fill distance over the adaptive SCMC, though CoMinED has smaller feasible ratio. One intuitive explanation is that by the way of candidate augmentation, CoMinED only refines the lattice grid up to certain granularity such that further refinement would not yield candidate samples that add value to the final space-filling design construction, even though those samples are likely feasible. This shows that the CoMinED naturally comes with the heuristic for adaptive resource allocation between exploration and exploitation during candidate augmentation. 

After getting the feasible candidate samples, the next step is to construct the desired $n$-point design using the candidate set. In this paper, we consider constructing both maximin and MaxPro designs by the one-point-at-a-time greedy algorithm. Given that the greedy approach likely results in local optimum, we allow for 10 restarts to obtain the design with best measure. For maximin measure, the larger the better \parencite{johnson1990design}; and for the Maxpro measure, the smaller the better \parencite{joseph2015maxpro}. From Figure~\ref{fig:moto}, we can see that CoMinED outperforms adaptive SCMC on both maximin and MaxPro design construction using the candidate set, and the improvement is more significant when the total sample size is small. From both the candidates quality metrics (feasible ratio and fill distance) and the resulted design measures (maximin and MaxPro), additional samples bring minimal benefit for CoMinED as it already cover the feasible space well with $Q = 5$ presented in Figure~\ref{fig:mot-comined}. The robust performance of CoMinED under small sample size makes it the favorable option especially when the constraints are expensive to evaluate. \par

\begin{figure}[t!]
  \centering

  \begin{subfigure}{0.45\textwidth}
    \centering
    \includegraphics[width=0.9\textwidth]{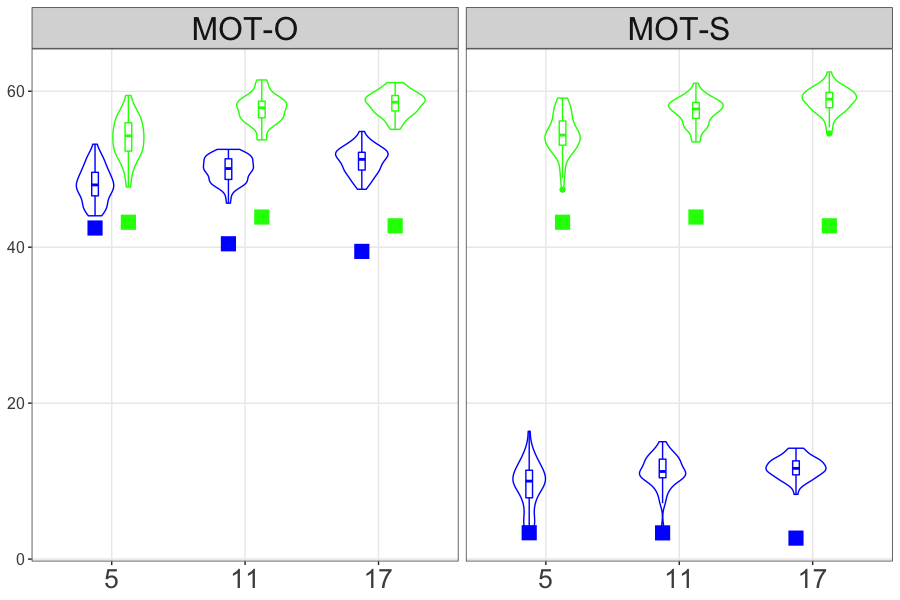}
    \caption{Feasible Ratio (Larger is Better)}
  \end{subfigure}%
  \begin{subfigure}{0.45\textwidth}
    \centering
    \includegraphics[width=0.9\textwidth]{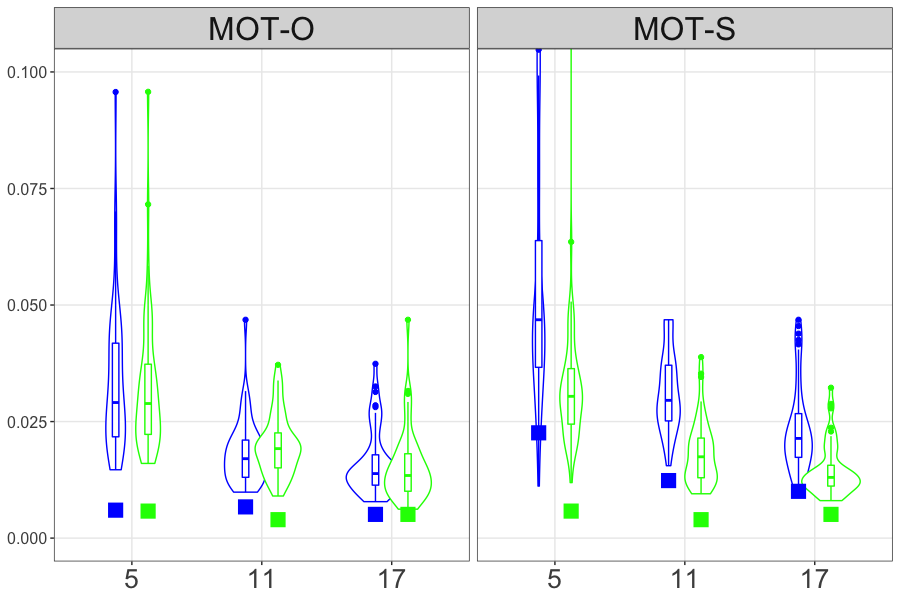}
    \caption{Fill Distance (Smaller is Better)}
  \end{subfigure}%

  \begin{subfigure}{0.45\textwidth}
    \centering
    \includegraphics[width=0.9\textwidth]{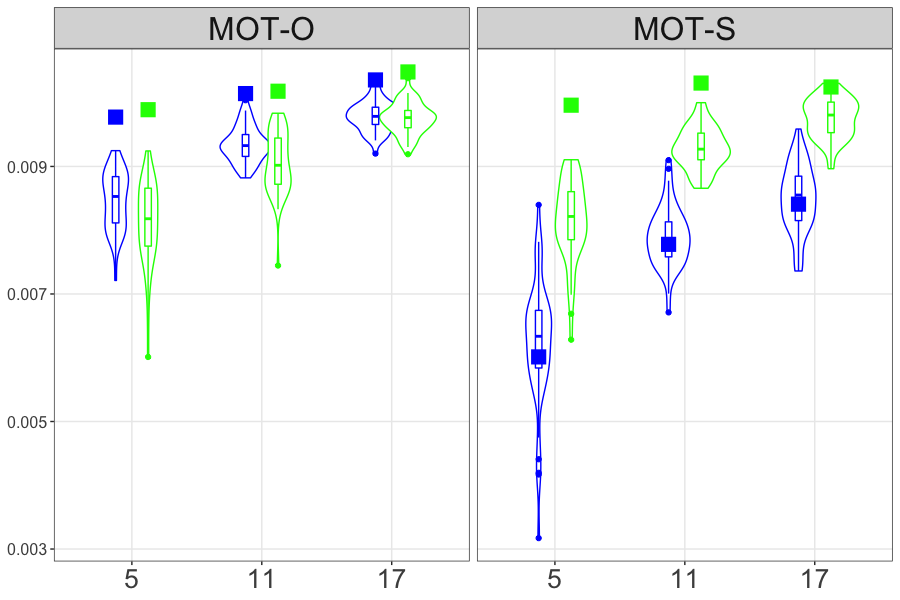}
    \caption{Maximin (Larger is Better)}
  \end{subfigure}%
  \begin{subfigure}{0.45\textwidth}
    \centering
    \includegraphics[width=0.9\textwidth]{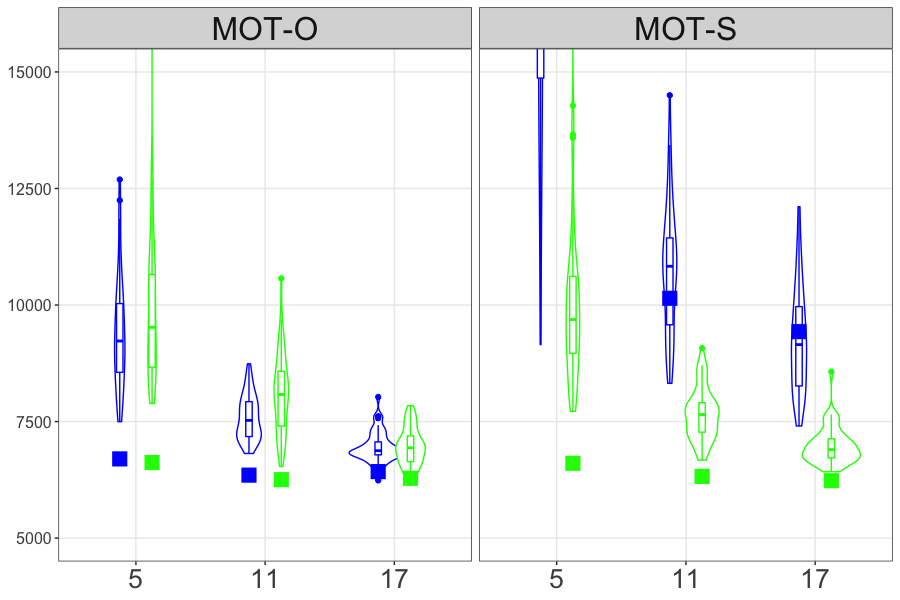}
    \caption{MaxPro (Smaller is Better)}
  \end{subfigure}%

  \caption{Comparisons of the candidates quality and the resulted 53-point design from applying CoMinED (squares) and adaptive SCMC (violin plots over 50 runs) on the motivation problem (MOT-O, \ref{eq:MOT}) and scaled motivation problem (MOT-S, \ref{eq:MOT-S}). CoMinED is ran with $Q=5,11,17$, and the corresponding adaptive SCMC comparison is ran with $M$ computed by \eqref{eq:scmc_M}. Results without (in blue) and with (in green) constraint value normalization are both presented.}
  \label{fig:mot_all}
\end{figure}

In many real world problems such as the welded beam design \parencite[WBD;][]{dong2018cgo} problem \eqref{eq:WBD}, it is common that the constraints yield values in very different scales, as $g_{3}$ of WBD is in the scale of 10s, but $g_{2}$ is in the scale of 10{,}000s. It is natural to test out how CoMinED and adaptive SCMC perform under the aforementioned circumstance. Consider the scaled version of the motivation example (MOT-S) presented below:
\begin{equation}
  \label{eq:MOT-S}
    \begin{aligned}
      &g_{1}(x) = 10^{-3}(x_{1} - \sqrt{50(x_{2}-0.52)^2+2} + 1) \leq 0 \\
      &g_{2}(x) = \sqrt{120(x_{2}-0.48)^2+1) - 0.75 - x_{1}} \leq 0 \\
      &g_{3}(x) = 10^{3}(0.65^2 - x_{1}^2 - x_{2}^2) \leq 0 \\
      \mbox{where} \quad & 0 \leq x_{i} \leq 1\; (i=1,2).
    \end{aligned}
\end{equation}
The blue squares and violin plots in Figure~\ref{fig:mot_all} are the results from applying CoMinED and adaptive SCMC directly on the motivation problem (MOT-O, \ref{eq:MOT}) and the scaled motivation problem (MOT-S, \ref{eq:MOT-S}). Comparing the MOT-S and the MOT-O facet, we can see that both CoMinED (blue squares) and adaptive SCMC (blue violin plots) perform substantially worse on all evaluation metrics for the scaled problem. Thus, we propose the \textit{constraint value normalization} using the median absolute deviation (MAD) centered at zero for remedy. Suppose that we have evaluated the constraints on some samples $\{x_{i}\}_{i=1}^{N^{\prime}}$, we replace the constraints $g_{k}$ in CoMinED and adaptive SCMC by 
\begin{equation}
  \label{eq:cvn}
    \tilde{g}_{k}(\cdot) = g_{k}(\cdot) / \sigma_{k}, \mbox{ where } \sigma_{k} = \mbox{Median}_{i=1:N^{\prime}}(|g_{k}(x_{i})-0|).
\end{equation}
The use of MAD instead of standard deviation is for robustness against large absolute constraint values. The green squares and violin plots in Figure~\ref{fig:mot_all} shows the performance of CoMinED and adaptive SCMC with the constraint value normalization. Compared to the results without the constraint value normalization (blue squares and violin plots), substantial improvements are observed, especially on the scaled motivation example. Hence, for the rest of the simulations, we only compare CoMinED to adaptive SCMC after incorporating constraint value normalization. 

\subsection{More Benchmark Problems}
\label{subsec:benchmark_problems}

\begin{figure}[t!]
  \centering
  \begin{subfigure}{0.45\textwidth}
    \centering
    \includegraphics[width=0.9\textwidth]{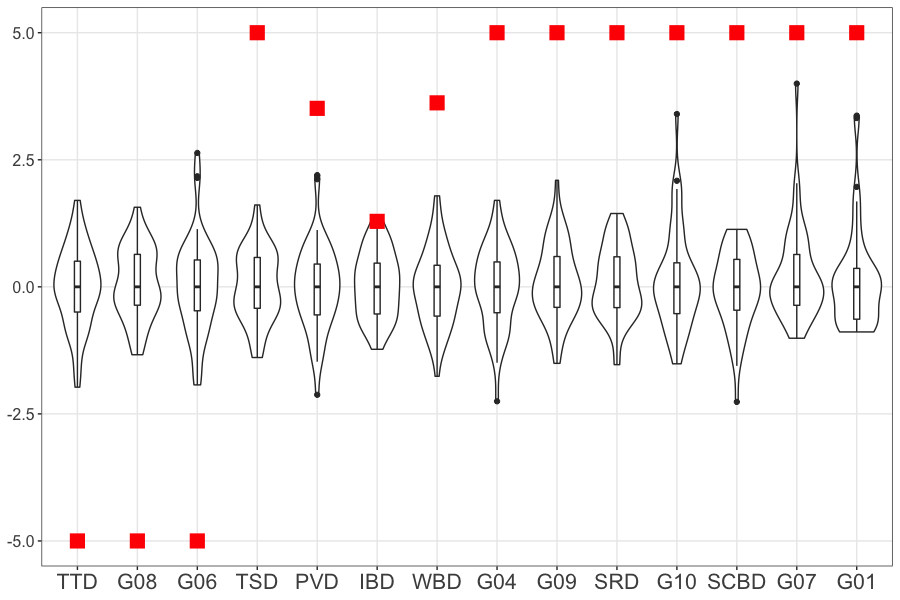}
    \caption{Feasible Ratio (Larger is Better)}
  \end{subfigure}%
  \begin{subfigure}{0.45\textwidth}
    \centering
    \includegraphics[width=0.9\textwidth]{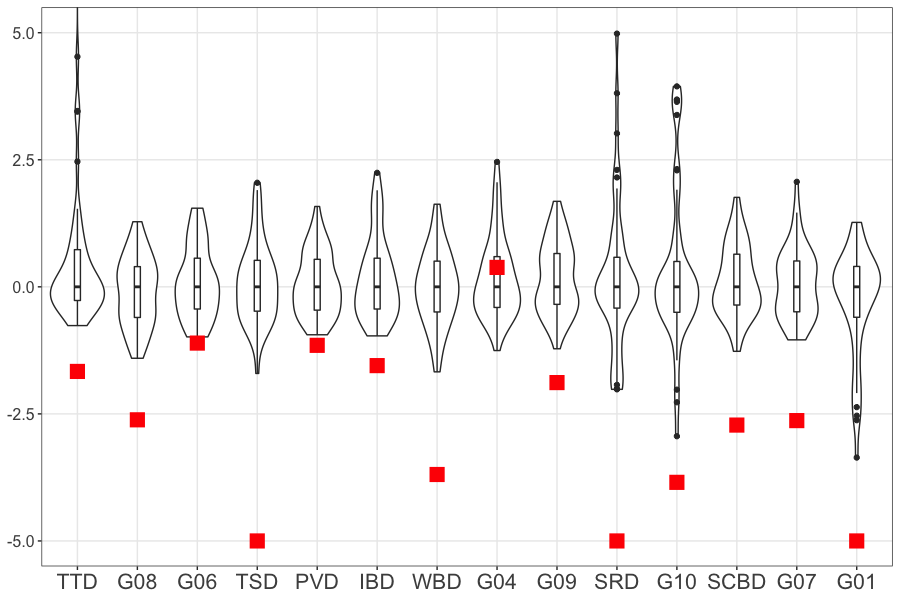}
    \caption{Fill Distance (Smaller is Better)}
  \end{subfigure}%

  \begin{subfigure}{0.45\textwidth}
    \centering
    \includegraphics[width=0.9\textwidth]{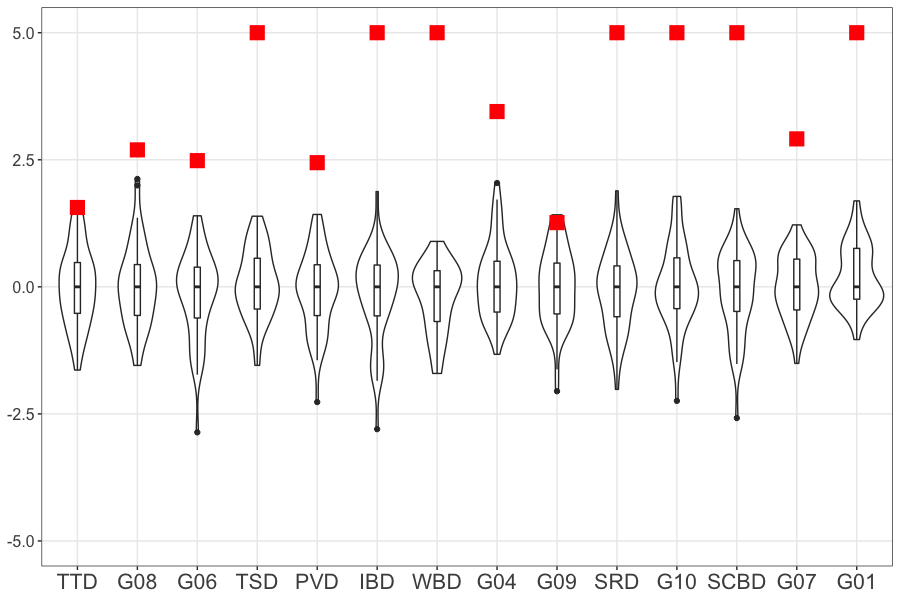}
    \caption{Maximin (Larger is Better)}
  \end{subfigure}%
  \begin{subfigure}{0.45\textwidth}
    \centering
    \includegraphics[width=0.9\textwidth]{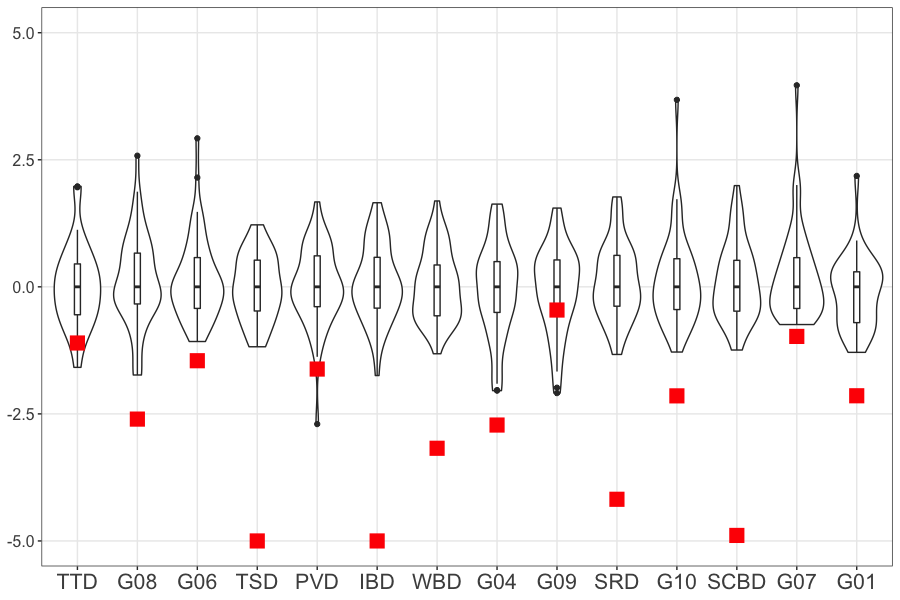}
    \caption{MaxPro (Smaller is Better)}
  \end{subfigure}%

  \caption{Comparisons of the candidates quality and the resulted 109-point design from applying CoMinED (squares) and adaptive SCMC (violin plots over 50 runs) on 14 benchmark problems. The problems are in ascending order by number of dimensions and descending order by the feasibility ratio. CoMinED is ran with $Q = 19$ for problems with dimension smaller than 10 and $Q = 27$ for problems with dimensions at least 10 (SCBD,G07,G01), and the corresponding adaptive SCMC comparison is ran with $M$ computed by \eqref{eq:scmc_M}. For each problem, the evaluation metrics are shifted such that the median of the adaptive SCMC results is 0 and re-scaled such that the IQR of adaptive SCMC results is 1. Both metrics are truncated at $\pm 5$ for visualization purpose.}

  \label{fig:benchmark_all}
\end{figure}

To further demonstrate the improvement of CoMinED, let us look at the simulation results on the 14 benchmark problems with dimensions ranging from 2 to 13, including 7 real world engineering problems. Top panels of Figure~\ref{fig:benchmark_all} compare the quality of the candidate samples. Except for the two-dimensional problems (TTD, G08, G06), CoMinED shows strong improvement in the feasible sample ratio, especially on the high dimensional problems. For example, 23.12\% of the total CoMinED samples are feasible for problem G01, while adaptive SCMC only yields in average 0.78\% feasible samples (See Table~\ref{tab:candidate_quality} in Appendix~\ref{appendix:simulation} for the actual feasible ratio value). On the other hand, looking at the fill distance\footnote{Fill distance is again approximated by $10^{4}$ feasible samples by acceptance/rejection sampling on a larger set of Sobol' points in unit hypercube. However, for problem G07, only $10^{3}$ samples are used since it is too expensive to generate more feasible samples by acceptance/rejection given its extremely small feasibility ratio (less than 1e-6).}, CoMinED outperforms the adaptive SCMC for all benchmark problems except G04, which is a 5 dimensional problem with feasibility ratio of 27.0\% (Table~\ref{tab:benchmark_info}). Given that G04 is an ``easy" problem, with more than 15{,}000 evaluations of the constraints (Table~\ref{tab:candidate_quality} in Appendix~\ref{appendix:simulation}), both CoMinED and adaptive SCMC should yield samples that cover the feasible region well. Next, let us look at the performance of the 109-point designs constructed from the feasible candidate samples. Similar to the motivation examples, we allow for 10 restarts of the one-point-at-a-time greedy algorithm to obtain the best design. From the bottom panels of Figure~\ref{fig:benchmark_all}, we can see that both maximin and MaxPro designs generated using the CoMinED candidates significantly outmatch the corresponding designs by the adaptive SCMC samples. In summary, from the extensive simulation results on the 14 benchmark problems, CoMinED is more robust than the adaptive SCMC for generating good space-filling design candidate points in arbitrary non-rectangular bounded design space, especially when the space is high dimensional and highly constrained.

\section{Conclusion}
\label{sec:conclusion}
This paper proposes the Constrained Minimum Energy Design (CoMinED; Algorithm~\ref{algo:comined}) for constructing space-filling designs in any non-rectangular bounded space defined by inequality constraints. The key idea of the CoMinED is to employ the state-of-the-art deterministic sampling algorithm, Minimum Energy Design (MinED), on the target distribution using the the probabilistic constraints proposed in Sequentially Constrained Monte Carlo (SCMC). Different from the use of local maximin LHDs for candidate augmentation in the MinED algorithm, we propose the adaptive lattice grid refinement that would impose restriction on the minimal interpoint spacing for the candidate samples, making them more favorable as the candidate set for space-filling designs construction. The extensive simulations on the 15 benchmark problems with dimensions ranging from 2 to 13 demonstrate the significant improvement of CoMinED over adaptive SCMC, the best candidate generation approach we can find from the existing literature. CoMinED also enjoys from fewer number of constraint evaluations by avoiding the sampling of the arbitrarily close points that add minimal value for the space-filling designs construction. However, many real world applications involve discrete variables, but CoMinED and adaptive SCMC can only handle continuous variables. One future research direction is to investigate how to construct constrained space-filling design on the set of mixed discrete and continuous variables. On the other hand, as some recent papers have proposed that ``sampling can be faster than optimization" \parencite{ma2019sampling}, another future research direction is to investigate how CoMinED might be useful for solving expensive constrained optimization problem.

\bigskip
\section*{\Large{Acknowledgments}}
This research is supported by a U.S. Army Research Office grant W911NF-17-1-0007.

\bigskip
\printbibliography

\bigskip

\section*{\LARGE{Appendix}}

\appendix

\section{Implementation Details of the CoMinED Algorithm}
\label{appendix:comined_details}
In this section we discuss some implementation details of CoMinED (Algorithm~\ref{algo:comined}) regarding the choice of parameters and ways to improve the computational efficiency. Let us assume that the number of design points $n$ and the number of nearest neighbors considered in candidate augmentation $Q$ are user specific. For $Q$, we suggest the use of any number between $2p+1$ and $3p+1$ depending on the available computational resource. One natural choice for $N_1$, the number of initial lattice candidate points, is the greatest prime number that is smaller than $Qn$. This approach is taken for all simulations ran in this paper. Next, let us discuss how to choose the increasing sequence of rigidity parameters $\{\tau_{t}\}_{t=0}^{T}$ with $\tau_{0} = 0$ and $\tau_{T} = 10^{6}$. In this paper, we use $\{\tau_{t}\}_{t=0}^{8} = [0,e^{1},e^{2},e^{3},e^{4},e^{5},e^{6},e^{7},10^{6}]$ that shows robust performance on problems with dimensions ranging from 2 to 13. If the user wants to try different number of intermediate steps $T$, one suggestion is to have them equally spaced in log-scale between $e^{0}$ and $e^{7}=1096.6$, i.e., 
\begin{equation}
  \label{eq:ridigity_para}
  \tau_{0} = 0, \tau_{T} = 10^{6}, \mbox{ and } \tau_{t} = \exp\bigg\{\frac{7}{T-1}t\bigg\} \mbox{ for } t = 1,\ldots,T-1.
\end{equation}
Moreover, as pointed out in \textcite{golchi2015scmc}, one adaptive approach to determine $\tau_{t+1}$ from the existing samples is to ensure that the effective sample size does not fall below certain threshold when we transition from $\tau_{t}$ to $\tau_{t+1}$. Similar idea is also applicable for CoMinED. \par

Last, we address one computational burden of CoMinED resulted from the increasing number of candidate samples used at the one-point-at-a-time greedy algorithm in the intermediate designs construction step as the algorithm proceeds. Recall our goal is to solve
\begin{equation}
  \label{eq:ap1_1}
    \mathcal{D}^{*}_n = \arg\max_{\mathcal{D}_n\subseteq\mathcal{C}^{t}} h(\mathcal{D}_n) = \min_{\substack{x_i,x_j\in\mathcal{D}_n\\i\neq j}}\frac{1}{2p}\log\rho_{\tau_{t}}(x_i) + \frac{1}{2p}\log\rho_{\tau_{t}}(x_j) + \log \lVert x_i - x_j\rVert_{s}\; ,
\end{equation}
where $\rho_{\tau_{t}}(\cdot) = \prod_{k=1}^{K}\Phi(-\tau_{t} g_{k}(\cdot))$ is the probabilistic constraints function and $\mathcal{C}^{t}$ is the $t$-th step $N_{t}$-point candidate set. One easy solution is to ignore samples in $\mathcal{C}^{t}$ that are not important, i.e., we only apply the greedy algorithm on $\{y\in\mathcal{C}^{t}:\log\rho_{\tau_{t}}(y)>\eta\}$ where $\eta$ is some threshold indicating that whether a sample would have a impact on the solution of \eqref{eq:ap1_1}. For the distance measure $s=2$ considered in this paper, we know that the minimal interpoint distance of all candidate points is $\delta_{t} = \delta/2^{t}$ where $\delta$ is the minimal interpoint distance of the initial candidate set. Thus, $\log \lVert x_i - x_j\rVert_{2}$ in \eqref{eq:ap1_1} is lower bounded by $\log \delta_{t}$. For notation simplicity, denote the log-likelihood value by $\nu_{i} = \log\rho_{\tau_{t}}(x_i)$. Let $\mathcal{D}^{\prime}_{n}$ be the $n$ samples from the candidate set $\mathcal{C}^{t}$ with the top log-likelihood value, and thus we have
\begin{equation}
  \label{eq:ap1_2}
  h(\mathcal{D}^{*}_{n}) \geq h(\mathcal{D}^{\prime}_{n}) \geq \frac{1}{2p}\nu_{(n^{+})} + \frac{1}{2p}\nu_{(n^{+})} + \log \delta_{t}.
\end{equation}
where $\nu_{(n^{+})}$ is the $(n^{+} = N_{t} - n + 1)$-th order statistics of $\{\nu_{i}\}_{i=1}^{N_{t}}$. Thus, we can set 
\begin{equation}
  \label{eq:ap1_3}
  \eta = \nu_{(n^{+})} + 2p\log \delta_{t}.
\end{equation}
To avoid numerical comparison issue causing by machine round-off, we use 2.5 instead of 2 in \eqref{eq:ap1_3} in actual implementation.

\section{Benchmark Problems}
\label{appendix:benchmark_problems}
\begin{itemize}
  \item G01 \parencite{liu2017cgo}:
  {\footnotesize
  \begin{equation}
    \label{eq:G01}
      \begin{aligned}
        \min_{x} \quad &f(x) = 5\sum_{i=1}^{4}x_{i} - 5\sum_{i=1}^{4}x_{i}^2 - \sum_{i=5}^{13}x_{i} \\
        \mbox{subject to} \quad &g_{1}(x) = 2x_{1} + 2x_{2} + x_{10} + x_{11} - 10 \leq 0 \\
        &g_{2}(x) = 2x_{1} + 2x_{3} + x_{10} + x_{12} - 10 \leq 0 \\
        &g_{3}(x) = 2x_{2} + 2x_{3} + x_{11} + x_{12} - 10 \leq 0 \\
        &g_{4}(x) = -8x_{1} + x_{10} \leq 0 \\
        &g_{5}(x) = -8x_{2} + x_{11} \leq 0 \\
        &g_{6}(x) = -8x_{3} + x_{12} \leq 0 \\
        &g_{7}(x) = -2x_{4} - x_{5} + x_{10} \leq 0 \\
        &g_{8}(x) = -2x_{6} - x_{7} + x_{11} \leq 0 \\
        &g_{9}(x) = -2x_{8} - x_{9} + x_{12} \leq 0 \\
        \mbox{where} \quad & 0 \leq x_{i} \leq 1 \; (i=1,\ldots,9,13) \mbox{ and } 0 \leq x_{i} \leq 100 \; (i=10,11,12)
      \end{aligned}
  \end{equation}
  }
  
  \item G04 \parencite{liu2017cgo}:
  {\footnotesize
  \begin{equation}
    \label{eq:G04}
      \begin{aligned}
        \min_{x} \quad &f(x) = 5.3578547 x_{3}^2 + 0.8356891 x_{1}x_{5} + 37.293239 x_{1} - 40792.141 \\
        \mbox{subject to} \quad &g_{1}(x) = 85.334407 + 0.0056858 x_{2}x_{5} + 0.0006262 x_{1}x_{4} - 0.0022053x_{3}x_{5} - 92 \leq 0 \\
        &g_{2}(x) = -85.334407 - 0.0056858 x_{2}x_{5} - 0.0006262 x_{1}x_{4} + 0.0022053x_{3}x_{5} \leq 0 \\
        &g_{3}(x) = 80.51249 + 0.0071317 x_{2}x_{5} + 0.0029955 x_{1}x_{2} + 0.0021813 x_{3}^2 - 110 \leq 0 \\
        &g_{4}(x) = -80.51249 - 0.0071317 x_{2}x_{5} - 0.0029955 x_{1}x_{2} - 0.0021813 x_{3}^2 + 90 \leq 0 \\
        &g_{5}(x) = 9.300961 + 0.0047026 x_{3}x_{5} + 0.0012547 x_{1}x_{3} + 0.0019085 x_{3}x_{4} - 25 \leq 0 \\
        &g_{6}(x) = -9.300961 - 0.0047026 x_{3}x_{5} - 0.0012547 x_{1}x_{3} - 0.0019085 x_{3}x_{4} + 20 \leq 0 \\
        \mbox{where} \quad & 78 \leq x_{1} \leq 102, \; 33 \leq x_{2} \leq 45, \; \mbox{and } 27 \leq x_{i} \leq 45\; (i = 3,4,5)
      \end{aligned}
  \end{equation}
  }
  
  \item G06 \parencite{liu2017cgo}:
  {\footnotesize
  \begin{equation}
    \label{eq:G06}
      \begin{aligned}
        \min_{x} \quad &f(x) = (x_{1}-10)^3 + (x_{2}-20)^3 \\
        \mbox{subject to} \quad &g_{1}(x) = -(x_{1}-5)^2 - (x_{2}-5)^2 + 100 \leq 0 \\
        &g_{2}(x) = (x_{1}-6)^2 + (x_{2}-5)^2 - 82.81 \leq 0  \\
        \mbox{where} \quad & 13 \leq x_{1} \leq 100 \mbox{ and } 0 \leq x_{2} \leq 100
      \end{aligned}
  \end{equation}
  }
  
  \item G07 \parencite{liu2017cgo}:
  {\footnotesize
  \begin{equation}
    \label{eq:G07}
      \begin{aligned}
        \min_{x} \quad &f(x) = x_{1}^2 + x_{2}^2 + x_{1}x_{2} - 14x_{1} - 16x_{2} + (x_{3}-10)^2 + 4(x_{4}-5)^2 + (x_{5}-3)^2 +  \\
        &\qquad \quad  2(x_{6}-1)^2 + 5x_{7}^2 + 7(x_{8}-11)^2 + 2(x_{9}-10)^2 + (x_{10}-7)^2 + 45 \\
        \mbox{subject to} \quad &g_{1}(x) = -105 + 4x_{1} + 5x_{2} - 3x_{7} + 9x_{8} \leq 0 \\
        &g_{2}(x) = 10x_{1} - 8x_{2} - 17x_{7} + 2x_{8} \leq 0  \\
        &g_{3}(x) = -8x_{1} + 2x_{2} + 5x_{9} - 2x_{10} - 12 \leq 0 \\
        &g_{4}(x) = 3(x_{1}-2)^2 + 4(x_{2}-3)^2 + 2x_{3}^2 - 7x_{4} - 120 \leq 0 \\
        &g_{5}(x) = 5x_{1}^2 + 8x_{2} + (x_{3}-6)^2 - 2x_{4} - 40 \leq 0 \\
        &g_{6}(x) = x_{1}^2 + 2(x_{2}-2)^2 - 2x_{1}x_{2} + 14x_{5} - 6x_{6} \leq 0 \\
        &g_{7}(x) = 0.5(x_{1}-8)^2 + 2(x_{2}-4)^2 + 3x_{5}^2 - x_{6} - 30 \leq 0 \\
        &g_{8}(x) = -3x_{1} + 6x_{2} + 12(x_{9}-8)^2 - 7x_{10} \leq 0 \\
        \mbox{where} \quad & -10 \leq x_{i} \leq 10 \; (i = 1,\ldots,10)
      \end{aligned}
  \end{equation}
  }
  
  \item G08 \parencite{liu2017cgo}:
  {\footnotesize
  \begin{equation}
    \label{eq:G08}
      \begin{aligned}
        \min_{x} \quad &f(x) = \frac{\sin^3(2\pi x_{1})\sin(2\pi x_{2})}{x_{1}^3(x_{1}+x_{2})} \\
        \mbox{subject to} \quad &g_{1}(x) = x_{1}^2 - x_{2} + 1 \leq 0 \\
        &g_{2}(x) = 1 - x_{1} + (x_{2}-4)^2 \leq 0  \\
        \mbox{where} \quad & 0 \leq x_{i} \leq 10 \; (i = 1,2)
      \end{aligned}
  \end{equation}
  }
  
  \item G09 \parencite{liu2017cgo}:
  {\footnotesize
  \begin{equation}
    \label{eq:G09}
      \begin{aligned}
        \min_{x} \quad &f(x) = (x_{1}-10)^2 + 5(x_{2}-12)^2 + x_{3}^4 + 3(x_{4}-11)^2 + 10x_{5}^6 + \\
        &\qquad \quad 7x_{6}^2 + x_{7}^4 - 4x_{6}x_{7} - 10x_{6} - 8x_{7} \\
        \mbox{subject to} \quad &g_{1}(x) = -127 + 2x_{1}^2 + 3x_{2}^4 + x_{3} + 4x_{4}^2 + 5x_{5} \leq 0 \\
        &g_{2}(x) = -282 + 7x_{1} + 3x_{2} + 10x_{3}^2 + x_{4} - x_{5} \leq 0  \\
        &g_{3}(x) = -196 + 23x_{1} + x_{2}^2 + 6x_{6}^2 - 8x_{7} \leq 0 \\
        &g_{4}(x) = 4x_{1}^2 + x_{2}^2 - 3x_{1}x_{2} + 2x_{3}^2 + 5x_{6} - 11x_{7} \leq 0 \\
        \mbox{where} \quad & -10 \leq x_{i} \leq 10 \; (i = 1,\ldots,7)
      \end{aligned}
  \end{equation}
  }
  
  \item G10 \parencite{liu2017cgo}:
  {\footnotesize
  \begin{equation}
    \label{eq:G10}
      \begin{aligned}
        \min_{x} \quad &f(x) = x_{1} + x_{2} + x_{3} \\
        \mbox{subject to} \quad &g_{1}(x) = -1 + 0.0025(x_{4}+x_{6}) \leq 0 \\
        &g_{2}(x) = -1 + 0.0025(x_{5}+x_{7}-x_{4}) \leq 0  \\
        &g_{3}(x) = -1 + 0.01(x_{8}-x_{5}) \leq 0 \\
        &g_{4}(x) = -x_{1}x_{6} + 833.33252x_{4} + 100x_{1} - 83333.333 \leq 0 \\
        &g_{5}(x) = -x_{2}x_{7} + 1250x_{5} + x_{2}x_{4} - 1250x_{4} \leq 0 \\
        &g_{6}(x) = -x_{3}x_{8} + 1250000 + x_{3}x_{5} - 2500x_{5} \leq 0 \\
        \mbox{where} \quad & 100 \leq x_{1} \leq 10000, \; 1000 \leq x_{i} \leq 10000\; (i=2,3), \; 10 \leq x_{i} \leq 1000\; (i=4,\ldots,8)
      \end{aligned}
  \end{equation}
  }
  
  \item I-Beam Design \parencite[IBD;][]{wang2003rslhd}:
  {\footnotesize
  \begin{equation}
    \label{eq:IBD}
      \begin{aligned}
        \min_{x} \quad &f(x) = \frac{5000}{\frac{x_{3}(x_{1}-2x_{4})^3}{12} + \frac{x_{2}x_{4}^3}{6} + 2 x_{2}x_{4}(\frac{x_{1}-x_{4}}{2})^2} \\
        \mbox{subject to} \quad &g_{1}(x) = 2 x_{2} x_{4} + x_{3}(x_{1}-2 x_{4}) - 300 \leq 0 \\
        &g_{2}(x) = \frac{180000 x_{1}}{x_{3}(x_{1}-2x_{4})^3 + 2x_{2}x_{4}[4x_{4}^2+3x_{1}(x_{1}-2x_{4})]} + \\
        &\qquad \quad \frac{15000 x_{2}}{(x_{1}-2x_{4})x_{3}^3+2x_{4}x_{2}^3} - 6 \leq 0\\
        \mbox{where} \quad & 10 \leq x_{1} \leq 80,\; 10 \leq x_{2} \leq 50,\; \mbox{and } 0.9 \leq x_{i} \leq 5\; (i=3,4)
      \end{aligned}
  \end{equation}
  }
  
  \item Pressure Vessel Design \parencite[PVD;][]{dong2018cgo,chaiyotha2020cgo}:
  {\footnotesize
  \begin{equation}
    \label{eq:PVD}
      \begin{aligned}
        \min_{x} \quad &f(x) = 0.6224 x_{1}x_{3}x_{4} +1.7781 x_{2}x_{3}^2 + 3.1661x_{1}^2x_{4} + 19.84x_{1}^2x_{3} \\
        \mbox{subject to} \quad &g_{1}(x) = -x_{1} + 0.0193x_{3} \leq 0 \\
        &g_{2}(x) = -x_{2} + 0.00954x_{3} \leq 0  \\
        &g_{3}(x) = -\pi x_{3}^2 x_{4} - \frac{4}{3} \pi x_{3}^3 + 1296000 \leq 0 \\
        &g_{4}(x) = x_{4} - 240 \leq 0 \\
        \mbox{where} \quad & 0.0625 \leq x_{i} \leq 6.1875\; (i=1,2) \mbox{ and } 10 \leq x_{i} \leq 200\; (i = 3,4)
      \end{aligned}
  \end{equation}
  }
  
  \item NASA Speed Reducer Design \parencite[SRD;][]{liu2017cgo,chaiyotha2020cgo}:
  {\footnotesize
  \begin{equation}
    \label{eq:SRD}
      \begin{aligned}
        \min_{x} \quad &f(x) = 0.7854 x_{1} x_{2}^2 (3.3333 x_{3}^2 + 14.9334 x_{3} - 43.0934) - 1.508 x_{1} (x_{6}^2 + x_{7}^2) + \\
        &\qquad \quad 7.4777(x_{6}^3+x_{7}^3) + 0.7854 (x_{4}x_{6}^2 + x_{5}x_{7}^2) \\
        \mbox{subject to} \quad &g_{1}(x) = \frac{27}{x_{1} x_{2}^2 x_{3}} - 1 \leq 0 \\
        &g_{2}(x) = \frac{397.5}{x_{1}x_{2}^2 x_{3}^2} - 1 \leq 0  \\
        &g_{3}(x) = \frac{1.93 x_{4}^3}{x_{2}x_{3}x_{6}^4} - 1 \leq 0 \\
        &g_{4}(x) = \frac{1.93 x_{5}^3}{x_{2}x_{3}x_{7}^4} - 1 \leq 0 \\
        &g_{5}(x) = \frac{\{(\frac{745 x_{4}}{x_{2}x_{3}})^2 + 16.9 \times 10^{6}\}^{0.5}}{110 x_{6}^3} - 1 \leq 0 \\
        &g_{6}(x) = \frac{\{(\frac{745 x_{5}}{x_{2}x_{3}})^2 + 157.5 \times 10^{6}\}^{0.5}}{85 x_{7}^3} - 1 \leq 0 \\
        &g_{7}(x) = \frac{x_{2}x_{3}}{40} - 1 \leq 0 \\
        &g_{8}(x) = \frac{5 x_{2}}{x_{1}} - 1 \leq 0 \\
        &g_{9}(x) = \frac{x_{1}}{12x_{2}} - 1 \leq 0 \\
        &g_{10}(x) = \frac{1.5 x_{6}+1.9}{x_{4}} - 1 \leq 0 \\
        &g_{11}(x) = \frac{1.1 x_{7}+1.9}{x_{5}} - 1 \leq 0 \\
        \mbox{where} \quad & 2.6 \leq x_{1} \leq 3.6,\; 0.7 \leq x_{2} \leq 0.8,\; 17 \leq x_{3} \leq 28,\; 7.3 \leq x_{4} \leq 8.3,\; \\
        & 7.8 \leq x_{5} \leq 8.3,\; 2.9 \leq x_{6} \leq 3.9,\; \mbox{and } 5 \leq x_{7} \leq 5.5
      \end{aligned}
  \end{equation}
  }
  
  \item Tension/Compression Spring Design \parencite[TSD;][]{dong2018cgo}:
  {\footnotesize
  \begin{equation}
    \label{eq:TSD}
      \begin{aligned}
        \min_{x} \quad &f(x) = x_{1}^2 x_{2} (x_{3}+2) \\
        \mbox{subject to} \quad &g_{1}(x) = 1 - \frac{x_{2}^3 x_{3}}{71875 x_{1}^4} \leq 0 \\
        &g_{2}(x) = \frac{4x_{2}^2-x_{1}x_{2}}{12566x_{1}^3(x_{2}-x_{1})} + \frac{1}{5108x_{1}^2} - 1 \leq 0  \\
        &g_{3}(x) = 1 - \frac{140.45x_{1}}{x_{3}x_{2}^2} \leq 0 \\
        &g_{4}(x) = \frac{x_{1}+x_{2}}{1.5} - 1 \leq 0 \\
        \mbox{where} \quad & 0.05 \leq x_{1} \leq 2,\; 0.25 \leq x_{2} \leq 1.3,\; \mbox{and } 2 \leq x_{3} \leq 15
      \end{aligned}
  \end{equation}
  }
  
  \item Three-Bar Truss Design \parencite[TTD;][]{liu2017cgo}:
  {\footnotesize
  \begin{equation}
    \label{eq:TTD}
      \begin{aligned}
        \min_{x} \quad &f(x) = (2\sqrt{2} x_{1} + x_{2}) \times l \\
        \mbox{subject to} \quad &g_{1}(x) = \frac{\sqrt{2}x_{1} + x_{2}}{\sqrt{2}x_{1}^2+2x_{1}x_{2}}P - \sigma \leq 0 \\
        &g_{2}(x) = \frac{x_{2}}{\sqrt{2}x_{1}^2+2x_{1}x_{2}}P - \sigma \leq 0  \\
        &g_{3}(x) = \frac{1}{x_{1} + \sqrt{2}x_{2}} P - \sigma \leq 0 \\
        \mbox{where} \quad & 0 \leq x_{i} \leq 1\; (i = 1,2),\; l = 100,\; P = 2,\; \mbox{and } \sigma = 2
      \end{aligned}
  \end{equation}
  }
  
  \item Welded Beam Design \parencite[WBD;][]{dong2018cgo} with modified ranges:
  {\footnotesize
  \begin{equation}
    \label{eq:WBD}
      \begin{aligned}
        \min_{x} \quad &f(x) = 1.10471 x_{1}^2 x_{2} + 0.04811 x_{3} x_{4} (14.0 + x_{2}) \\
        \mbox{subject to} \quad &g_{1}(x) = \tau - \tau_{\text{max}} \leq 0 \\
        &g_{2}(x) = \sigma - \sigma_{\text{max}} \leq 0  \\
        &g_{3}(x) = x_{1} - x_{4} \leq 0 \\
        &g_{4}(x) = 0.10471x_{1}^2 + 0.04811x_{3}x_{4}(14+x_{2}) - 5 \leq 0 \\
        &g_{5}(x) = \delta - \delta_{\text{max}} \leq 0 \\
        &g_{6}(x) = P - P_{c} \leq 0 \\
        \mbox{where} \quad & 0.125 \leq x_{1} \leq 10 \mbox{ and } 0.1 \leq x_{i} \leq 10\; (i = 2,3,4) \\
        & P = 6000,\; L = 14,\; E = 30\times 10^{6},\; G = 12\times 10^{6}, \\
        & \tau_{\text{max}} = 13600,\; \sigma_{\text{max}} = 30000,\; \delta_{\text{max}} = 0.25 \\
        & M = P (L + x_{2} / 2) \\
        & R = \sqrt{x_{2}^2/4 + (x_{1}+x_{3})^2/4} \\
        & J = 2\sqrt{2} x_{1}x_{2}(x_{2}^2/12+(x_{1}+x_{3})^2/4) \\
        & \tau_{1} = P/(\sqrt{2}x_{1}x_{2}) \\
        & \tau_{2} = MR/J \\
        & \tau = \sqrt{\tau_{1}^2 + 2\tau_{1}\tau_{2}\frac{x_{2}}{2R} + \tau_{2}^2} \\
        & \sigma = 6PL / (x_{4}x_{3}^2) \\
        & \delta = 4PL^3 / (E x_{3}^3 x_{4}) \\
        & P_{c} = \frac{4.013 E \sqrt{x_{3}^2 x_{4}^6/36}}{L^2}\bigg(1 - \frac{x_{3}}{2L}\sqrt{\frac{E}{4G}}\bigg) \\
      \end{aligned}
  \end{equation}
  }
  
  \newpage
  \item Stepped Cantilever Beam Design \parencite[SCBD;][]{dong2018cgo} with slight modification based on a \href{https://fr.mathworks.com/help/gads/solving-a-mixed-integer-engineering-design-problem-using-the-genetic-algorithm.html}{MathWorks documentation}:
  {\footnotesize
  \begin{equation}
    \label{eq:SCBD}
      \begin{aligned}
        \min_{b,h} \quad &f(b,h) = l\sum_{i=1}^{5}b_{i}h_{i} \\
        \mbox{subject to} \quad &g_{1}(b,h) = \frac{6P\cdot l}{b_{5}h_{5}^2} - 14000 \leq 0 \\
        &g_{2}(b,h) = \frac{6P\cdot 2l}{b_{4}h_{4}^2} - 14000 \leq 0  \\
        &g_{3}(b,h) = \frac{6P\cdot 3l}{b_{3}h_{3}^2} - 14000 \leq 0 \\
        &g_{4}(b,h) = \frac{6P\cdot 4l}{b_{2}h_{2}^2} - 14000 \leq 0 \\
        &g_{5}(b,h) = \frac{6P\cdot 5l}{b_{1}h_{1}^2} - 14000 \leq 0 \\
        &g_{6}(b,h) = \frac{P l^3}{3E}\bigg(\frac{61}{I_{1}}+\frac{37}{I_{2}}+\frac{19}{I_{3}}+\frac{7}{I_{4}}+\frac{1}{I_{5}}\bigg) - 2.7 \leq 0 \\
        &g_{7}(b,h) = \frac{h_{1}}{b_{1}} - 20 \leq 0 \\
        &g_{8}(b,h) = \frac{h_{2}}{b_{2}} - 20 \leq 0  \\
        &g_{9}(b,h) = \frac{h_{3}}{b_{3}} - 20 \leq 0 \\
        &g_{10}(b,h) = \frac{h_{4}}{b_{4}} - 20 \leq 0 \\
        &g_{11}(b,h) = \frac{h_{5}}{b_{5}} - 20 \leq 0 \\
        \mbox{where} \quad & 2 \leq b_{i} \leq 3.5\; (i = 1,\ldots,5) \mbox{ and } 35 \leq h_{i} \leq 60\; (i = 1,\ldots,5) \\
        & l = 100,\; P = 50000,\; E = 2\times 10^{7}\; \\
        & I_{i} = b_{i}h_{i}^{3}/12\; (i = 1,\ldots,5) \\
      \end{aligned}
  \end{equation}
  }
\end{itemize}

\section{Additional Simulation Results}
\label{appendix:simulation}

\begin{figure}[t!]
  \centering
  \begin{subfigure}{0.24\textwidth}
    \centering
    \includegraphics[width=0.95\textwidth]{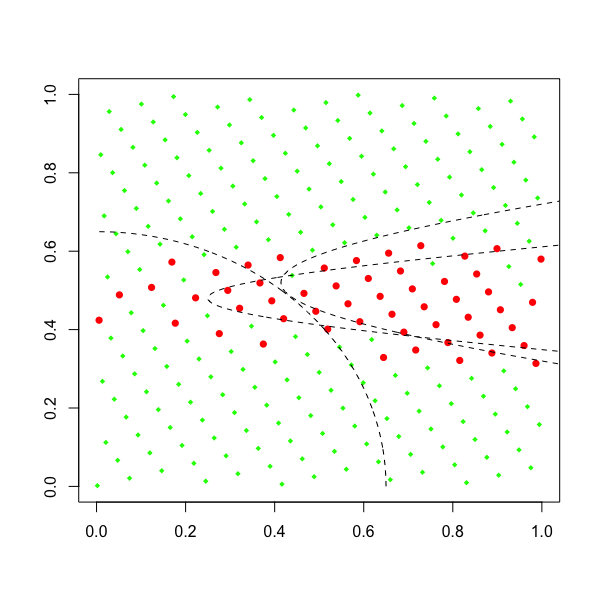}
    \caption{$t=1$}
  \end{subfigure}%
  \begin{subfigure}{0.24\textwidth}
    \centering
    \includegraphics[width=0.95\textwidth]{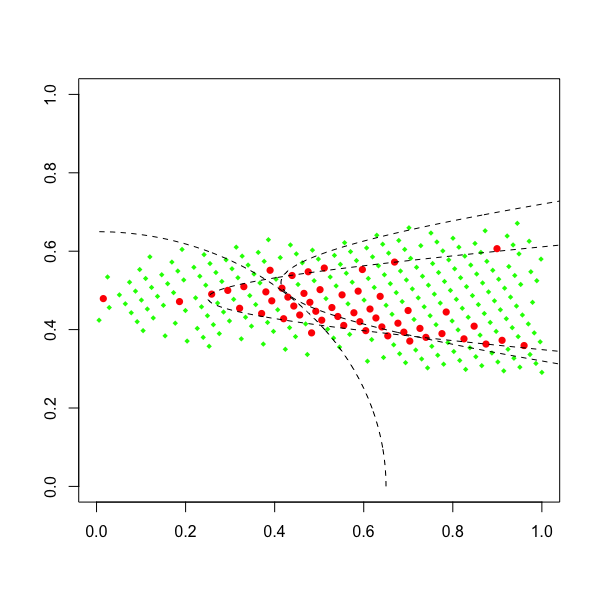}
    \caption{$t=2$}
  \end{subfigure}%
  \begin{subfigure}{0.24\textwidth}
    \centering
    \includegraphics[width=0.95\textwidth]{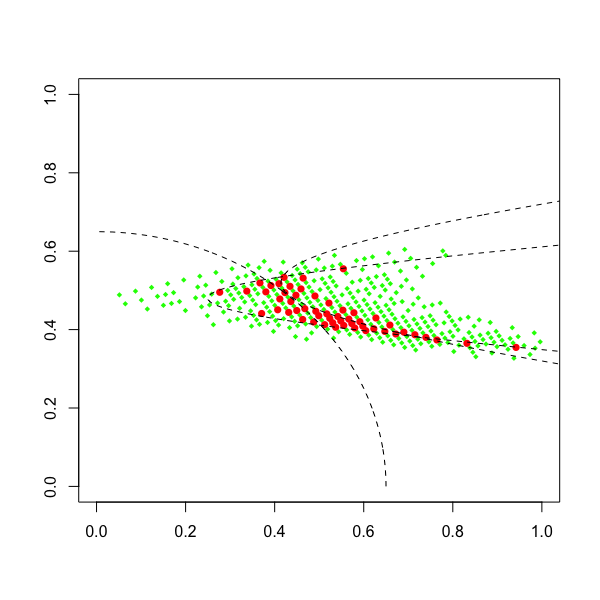}
    \caption{$t=3$}
  \end{subfigure}%
  \begin{subfigure}{0.24\textwidth}
    \centering
    \includegraphics[width=0.95\textwidth]{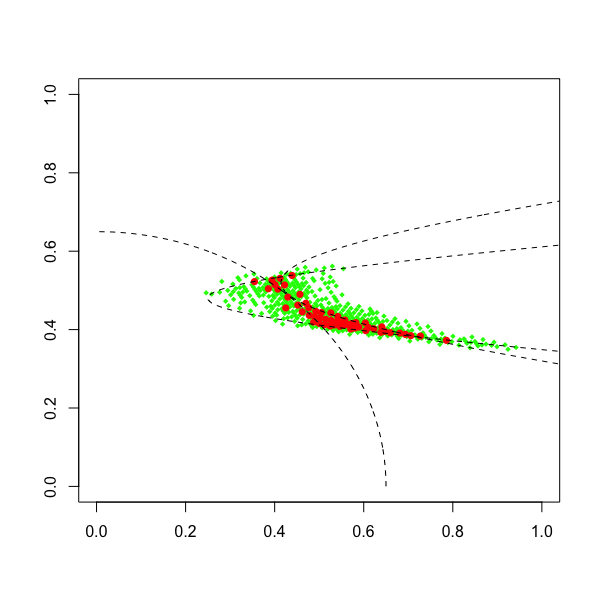}
    \caption{$t=4$}
  \end{subfigure}%
  
  \begin{subfigure}{0.24\textwidth}
    \centering
    \includegraphics[width=0.95\textwidth]{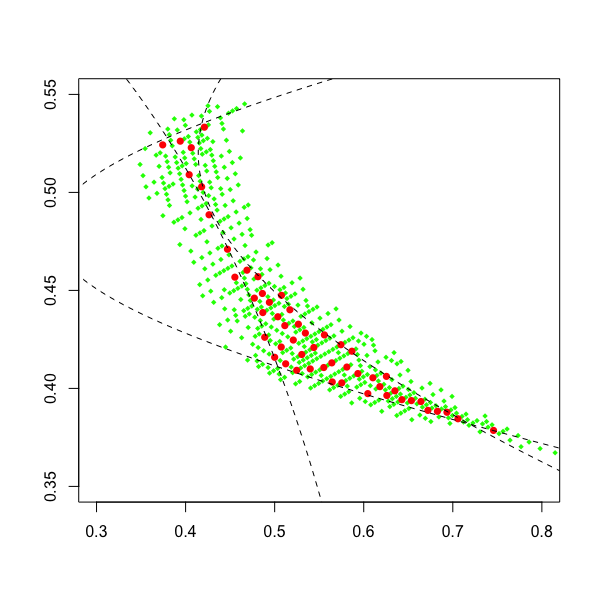}
    \caption{$t=5$}
  \end{subfigure}%
  \begin{subfigure}{0.24\textwidth}
    \centering
    \includegraphics[width=0.95\textwidth]{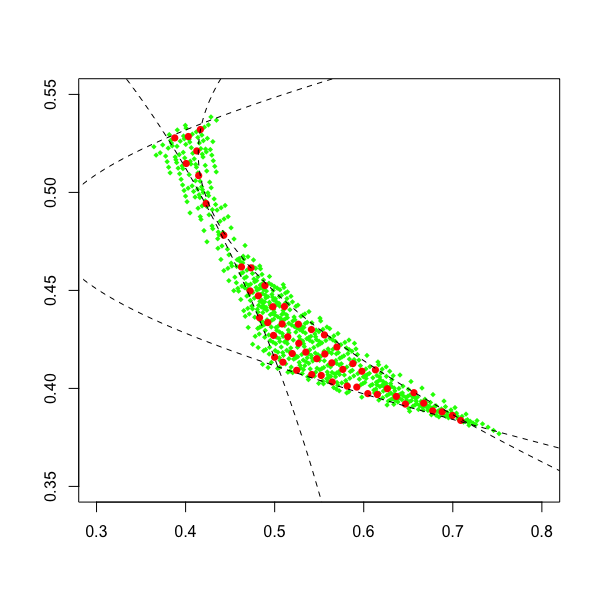}
    \caption{$t=6$}
  \end{subfigure}%
  \begin{subfigure}{0.24\textwidth}
    \centering
    \includegraphics[width=0.95\textwidth]{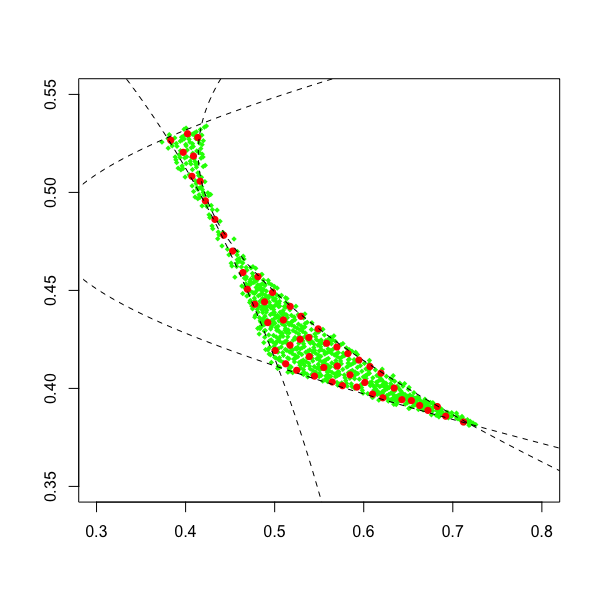}
    \caption{$t=7$}
  \end{subfigure}%
  \begin{subfigure}{0.24\textwidth}
    \centering
    \includegraphics[width=0.95\textwidth]{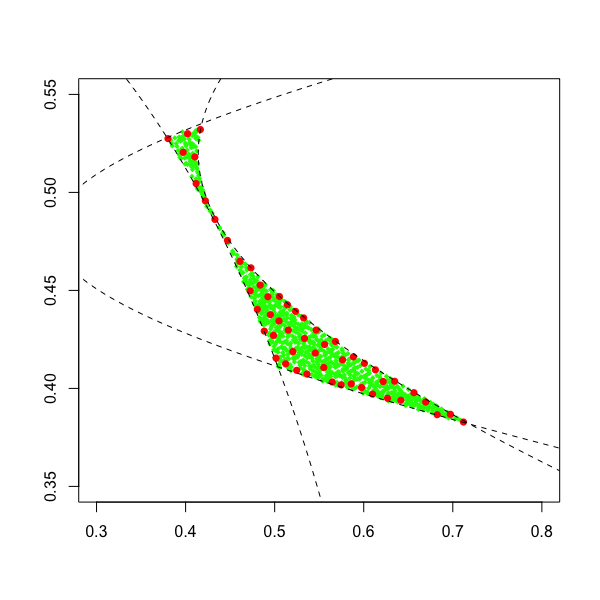}
    \caption{$t=8$}
  \end{subfigure}%
  
  \caption{Evolution of the CoMinED algorithm on the motivation problem \eqref{eq:MOT} with $n = 53$ points design, $Q = 5$, $N_{1} = 263$, $T = 8$, and $\{\tau_{t}\}_{t=0}^{8} = [0,e^{1},e^{2},e^{3},e^{4},e^{5},e^{6},e^{7},10^{6}]$. The green diamonds indicate the important candidate samples at each step (see Appendix~\ref{appendix:comined_details} for details), and the red circles indicate the intermediate 53-point CoMinEDs.}
  \label{fig:mot-comined_evolution}
\end{figure}

\begin{table}[t!]
  \centering
  \resizebox{\columnwidth}{!}{%
  \begin{tabular}{|ccc|cc|ccc|cccc|cccc|}
  \hline
  \multicolumn{3}{|c|}{} & \multicolumn{2}{c|}{} & \multicolumn{3}{c|}{LHDs} & \multicolumn{4}{c|}{adaptive SCMC} & \multicolumn{4}{c|}{CoMinED} \\
   Problem & $p$ & Feas. Ratio & $n$ & CVN & No. Cand. & Feas. Ratio & Fill Dist. & $M$ & No. Cand. & Feas. Ratio & Fill Dist. & $Q$ & No. Cand. & Feas. Ratio & Fill Dist. \\ 
   \hline
MOT-O & 2 & 0.53\% & 53 & No & 2155 & 0.53\% & 9.04e-02 & 265 & 2385 & \textcolor{red}{48.04\%} & 3.36e-02 & 5 & 2155 & 42.46\% & \textcolor{red}{6.00e-03} \\ 
MOT-O & 2 & 0.53\% & 53 & Yes & 1993 & 0.55\% & 9.41e-02 & 265 & 2385 & \textcolor{red}{54.01\%} & 3.23e-02 & 5 & 1993 & 43.20\% & \textcolor{red}{5.79e-03} \\ 
MOT-O & 2 & 0.53\% & 53 & No & 4951 & 0.55\% & 5.63e-02 & 583 & 5247 & \textcolor{red}{49.99\%} & 1.83e-02 & 11 & 4951 & 40.44\% & \textcolor{red}{6.70e-03} \\ 
MOT-O & 2 & 0.53\% & 53 & Yes & 4477 & 0.56\% & 5.70e-02 & 583 & 5247 & \textcolor{red}{57.67\%} & 1.99e-02 & 11 & 4477 & 43.87\% & \textcolor{red}{3.96e-03} \\ 
MOT-O & 2 & 0.53\% & 53 & No & 7689 & 0.52\% & 4.07e-02 & 901 & 8109 & \textcolor{red}{51.14\%} & 1.60e-02 & 17 & 7689 & 39.46\% & \textcolor{red}{5.09e-03} \\ 
MOT-O & 2 & 0.53\% & 53 & Yes & 7117 & 0.53\% & 4.95e-02 & 901 & 8109 & \textcolor{red}{58.44\%} & 1.53e-02 & 17 & 7117 & 42.74\% & \textcolor{red}{5.09e-03} \\ 
\hdashline
MOT-S & 2 & 0.53\% & 53 & No & 2338 & 0.53\% & 8.46e-02 & 265 & 2385 & \textcolor{red}{9.57\%} & 5.44e-02 & 5 & 2338 & 3.38\% & \textcolor{red}{2.26e-02} \\ 
MOT-S & 2 & 0.53\% & 53 & Yes & 1993 & 0.54\% & 9.92e-02 & 265 & 2385 & \textcolor{red}{54.46\%} & 3.30e-02 & 5 & 1993 & 43.20\% & \textcolor{red}{5.79e-03} \\ 
MOT-S & 2 & 0.53\% & 53 & No & 5146 & 0.53\% & 5.48e-02 & 583 & 5247 & \textcolor{red}{11.24\%} & 3.16e-02 & 11 & 5146 & 3.36\% & \textcolor{red}{1.23e-02} \\ 
MOT-S & 2 & 0.53\% & 53 & Yes & 4477 & 0.54\% & 6.18e-02 & 583 & 5247 & \textcolor{red}{57.45\%} & 1.82e-02 & 11 & 4477 & 43.87\% & \textcolor{red}{3.96e-03} \\ 
MOT-S & 2 & 0.53\% & 53 & No & 7952 & 0.52\% & 4.29e-02 & 901 & 8109 & \textcolor{red}{11.72\%} & 2.35e-02 & 17 & 7952 & 2.69\% & \textcolor{red}{1.00e-02} \\ 
MOT-S & 2 & 0.53\% & 53 & Yes & 7117 & 0.52\% & 4.90e-02 & 901 & 8109 & \textcolor{red}{58.87\%} & 1.45e-02 & 17 & 7117 & 42.74\% & \textcolor{red}{5.09e-03} \\
\hdashline
TTD & 2 & 21.79\% & 109 & Yes & 18504 & 21.81\% & 1.40e-02 & 2071 & 18639 & \textcolor{red}{85.07\%} & 1.79e-02 & 19 & 18504 & 82.91\% & \textcolor{red}{1.07e-02} \\ 
G08 & 2 & 0.86\% & 109 & Yes & 15111 & 0.87\% & 1.58e-02 & 2071 & 18639 & \textcolor{red}{68.79\%} & 7.29e-03 & 19 & 15111 & 63.59\% & \textcolor{red}{2.15e-03} \\ 
G06 & 2 & 0.01\% & 109 & Yes & 12579 & 0.01\% & 4.71e-02 & 2071 & 18639 & \textcolor{red}{28.45\%} & 8.62e-03 & 19 & 12579 & 11.69\% & \textcolor{red}{1.01e-03} \\ 
TSD & 3 & 0.75\% & 109 & Yes & 17725 & 0.75\% & 1.87e-01 & 2071 & 18639 & 40.51\% & 1.77e-01 & 19 & 17725 & \textcolor{red}{62.45\%} & \textcolor{red}{2.89e-02} \\ 
PVD & 4 & 40.32\% & 109 & Yes & 19025 & 40.34\% & 1.27e-01 & 2114 & 19026 & 72.55\% & 1.54e-01 & 19 & 19025 & \textcolor{red}{75.31\%} & \textcolor{red}{1.26e-01} \\ 
IBD & 4 & 0.15\% & 109 & Yes & 19292 & 0.16\% & 3.31e-01 & 2144 & 19296 & 43.82\% & 1.87e-01 & 19 & 19292 & \textcolor{red}{48.12\%} & \textcolor{red}{6.57e-02} \\ 
WBD & 4 & 0.10\% & 109 & Yes & 20858 & 0.10\% & 3.29e-01 & 2318 & 20862 & 35.96\% & 1.86e-01 & 19 & 20858 & \textcolor{red}{43.65\%} & \textcolor{red}{4.20e-02} \\ 
G04 & 5 & 26.96\% & 109 & Yes & 15467 & 26.92\% & 2.02e-01 & 2071 & 18639 & 60.32\% & \textcolor{red}{1.92e-01} & 19 & 15467 & \textcolor{red}{75.40\%} & 1.97e-01 \\ 
G09 & 7 & 0.53\% & 109 & Yes & 20845 & 0.52\% & 4.15e-01 & 2317 & 20853 & 32.88\% & 2.86e-01 & 19 & 20845 & \textcolor{red}{40.00\%} & \textcolor{red}{2.40e-01} \\ 
SRD & 7 & 0.19\% & 109 & Yes & 16219 & 0.18\% & 6.36e-01 & 2071 & 18639 & 18.19\% & 5.24e-01 & 19 & 16219 & \textcolor{red}{44.73\%} & \textcolor{red}{3.16e-01} \\ 
G10 & 8 & 0.00\% & 109 & Yes & 21438 & 0.00\% & 1.12e+00 & 2382 & 21438 & 6.45\% & 8.10e-01 & 19 & 21438 & \textcolor{red}{31.82\%} & \textcolor{red}{4.96e-01} \\ 
\hdashline
SCBD & 10 & 0.05\% & 109 & Yes & 21734 & 0.05\% & 8.77e-01 & 2943 & 26487 & 15.60\% & 6.81e-01 & 27 & 21734 & \textcolor{red}{48.37\%} & \textcolor{red}{4.90e-01} \\ 
G07 & 10 & 0.00\% & 109 & Yes & 29897 & 0.00\% & 1.11e+00 & 3322 & 29898 & 6.38\% & 6.47e-01 & 27 & 29897 & \textcolor{red}{29.36\%} & \textcolor{red}{3.82e-01} \\ 
G01 & 13 & 0.00\% & 109 & Yes & 22676 & 0.00\% & 2.01e+00 & 2943 & 26487 & 0.78\% & 1.86e+00 & 27 & 22676 & \textcolor{red}{23.12\%} & \textcolor{red}{1.38e+00} \\ 
  \hline
  \end{tabular}
  }
  \caption{Summary of simulation results on the feasible ratio (the larger the better) and the fill distance (the smaller the better) of the candidate set from applying one-step acceptance/rejection LHDs, adaptive SCMC, and CoMinED on all benchmark problems. The values for LHDs and adaptive SCMC are average over 50 runs. CVN stands for constraint value normalization.}
  \label{tab:candidate_quality}
\end{table}

\begin{table}[t!]
  \centering
  \resizebox{0.95\columnwidth}{!}{%
  \begin{tabular}{|ccc|cc|ccc|cccc|cccc|}
  \hline
  \multicolumn{3}{|c|}{} & \multicolumn{2}{c|}{} & \multicolumn{3}{c|}{LHDs} & \multicolumn{4}{c|}{adaptive SCMC} & \multicolumn{4}{c|}{CoMinED} \\
   Problem & $p$ & Feas. Ratio & $n$ & CVN & No. Cand. & Maximin & MaxPro & $M$ & No. Cand. & Maximin & MaxPro & $Q$ & No. Cand. & Maximin & MaxPro \\ 
   \hline
MOT-O & 2 & 0.53\% & 53 & No & 2155 & NaN & NaN & 265 & 2385 & 8.47e-03 & 9.41e+03 & 5 & 2155 & \textcolor{red}{9.77e-03} & \textcolor{red}{6.70e+03} \\ 
MOT-O & 2 & 0.53\% & 53 & Yes & 1993 & NaN & NaN & 265 & 2385 & 8.12e-03 & 9.94e+03 & 5 & 1993 & \textcolor{red}{9.89e-03} & \textcolor{red}{6.62e+03} \\ 
MOT-O & 2 & 0.53\% & 53 & No & 4951 & NaN & NaN & 583 & 5247 & 9.34e-03 & 7.58e+03 & 11 & 4951 & \textcolor{red}{1.01e-02} & \textcolor{red}{6.35e+03} \\ 
MOT-O & 2 & 0.53\% & 53 & Yes & 4477 & NaN & NaN & 583 & 5247 & 9.06e-03 & 8.10e+03 & 11 & 4477 & \textcolor{red}{1.02e-02} & \textcolor{red}{6.25e+03} \\ 
MOT-O & 2 & 0.53\% & 53 & No & 7689 & NaN & NaN & 901 & 8109 & 9.78e-03 & 6.94e+03 & 17 & 7689 & \textcolor{red}{1.04e-02} & \textcolor{red}{6.43e+03} \\ 
MOT-O & 2 & 0.53\% & 53 & Yes & 7117 & 1.43e-03 & 8.82e+04 & 901 & 8109 & 9.75e-03 & 6.96e+03 & 17 & 7117 & \textcolor{red}{1.05e-02} & \textcolor{red}{6.28e+03} \\ 
\hdashline
MOT-S & 2 & 0.53\% & 53 & No & 2338 & NaN & NaN & 265 & 2385 & \textcolor{red}{6.25e-03} & \textcolor{red}{1.89e+04} & 5 & 2338 & 6.01e-03 & 2.13e+04 \\ 
MOT-S & 2 & 0.53\% & 53 & Yes & 1993 & NaN & NaN & 265 & 2385 & 8.15e-03 & 1.00e+04 & 5 & 1993 & \textcolor{red}{9.96e-03} & \textcolor{red}{6.61e+03} \\ 
MOT-S & 2 & 0.53\% & 53 & No & 5146 & NaN & NaN & 583 & 5247 & \textcolor{red}{7.90e-03} & 1.08e+04 & 11 & 5146 & 7.78e-03 & \textcolor{red}{1.01e+04} \\ 
MOT-S & 2 & 0.53\% & 53 & Yes & 4477 & NaN & NaN & 583 & 5247 & 9.29e-03 & 7.66e+03 & 11 & 4477 & \textcolor{red}{1.03e-02} & \textcolor{red}{6.32e+03} \\ 
MOT-S & 2 & 0.53\% & 53 & No & 7952 & 3.26e-03 & 2.81e+04 & 901 & 8109 & \textcolor{red}{8.50e-03} & \textcolor{red}{9.27e+03} & 17 & 7952 & 8.41e-03 & 9.43e+03 \\ 
MOT-S & 2 & 0.53\% & 53 & Yes & 7117 & NaN & NaN & 901 & 8109 & 9.76e-03 & 6.98e+03 & 17 & 7117 & \textcolor{red}{1.02e-02} & \textcolor{red}{6.23e+03} \\ 
\hdashline
TTD & 2 & 21.79\% & 109 & Yes & 18504 & 3.84e-02 & 4.92e+02 & 2071 & 18639 & 3.92e-02 & 4.73e+02 & 19 & 18504 & \textcolor{red}{4.01e-02} & \textcolor{red}{4.68e+02} \\ 
G08 & 2 & 0.86\% & 109 & Yes & 15111 & 3.11e-03 & 4.47e+04 & 2071 & 18639 & 7.47e-03 & 1.14e+04 & 19 & 15111 & \textcolor{red}{7.76e-03} & \textcolor{red}{1.09e+04} \\ 
G06 & 2 & 0.01\% & 109 & Yes & 12579 & NaN & NaN & 2071 & 18639 & 6.67e-04 & 1.10e+06 & 19 & 12579 & \textcolor{red}{8.53e-04} & \textcolor{red}{7.06e+05} \\ 
TSD & 3 & 0.75\% & 109 & Yes & 17725 & 2.60e-02 & 5.80e+03 & 2071 & 18639 & 4.86e-02 & 1.86e+03 & 19 & 17725 & \textcolor{red}{7.09e-02} & \textcolor{red}{1.34e+03} \\ 
PVD & 4 & 40.32\% & 109 & Yes & 19025 & 2.66e-01 & 9.28e+01 & 2114 & 19026 & 2.64e-01 & 9.20e+01 & 19 & 19025 & \textcolor{red}{2.69e-01} & \textcolor{red}{9.05e+01} \\ 
IBD & 4 & 0.15\% & 109 & Yes & 19292 & NaN & NaN & 2144 & 19296 & 6.49e-02 & 1.04e+03 & 19 & 19292 & \textcolor{red}{7.47e-02} & \textcolor{red}{8.16e+02} \\ 
WBD & 4 & 0.10\% & 109 & Yes & 20858 & NaN & NaN & 2318 & 20862 & 6.18e-02 & 1.50e+03 & 19 & 20858 & \textcolor{red}{7.69e-02} & \textcolor{red}{1.20e+03} \\ 
G04 & 5 & 26.96\% & 109 & Yes & 15467 & 3.52e-01 & 8.52e+01 & 2071 & 18639 & 3.52e-01 & 8.26e+01 & 19 & 15467 & \textcolor{red}{3.62e-01} & \textcolor{red}{8.09e+01} \\ 
G09 & 7 & 0.53\% & 109 & Yes & 20845 & 1.49e-01 & 1.03e+03 & 2317 & 20853 & 2.95e-01 & 2.19e+02 & 19 & 20845 & \textcolor{red}{2.99e-01} & \textcolor{red}{2.18e+02} \\ 
SRD & 7 & 0.19\% & 109 & Yes & 16219 & NaN & NaN & 2071 & 18639 & 2.39e-01 & 4.03e+02 & 19 & 16219 & \textcolor{red}{3.13e-01} & \textcolor{red}{3.05e+02} \\ 
G10 & 8 & 0.00\% & 109 & Yes & 21438 & NaN & NaN & 2382 & 21438 & 8.52e-02 & 3.13e+03 & 19 & 21438 & \textcolor{red}{1.98e-01} & \textcolor{red}{7.17e+02} \\ 
\hdashline
SCBD & 10 & 0.05\% & 109 & Yes & 21734 & NaN & NaN & 2943 & 26487 & 3.38e-01 & 2.88e+02 & 27 & 21734 & \textcolor{red}{4.16e-01} & \textcolor{red}{2.11e+02} \\ 
G07 & 10 & 0.00\% & 109 & Yes & 29897 & NaN & NaN & 3322 & 29898 & 1.40e-01 & 1.91e+03 & 27 & 29897 & \textcolor{red}{2.43e-01} & \textcolor{red}{8.69e+02} \\ 
G01 & 13 & 0.00\% & 109 & Yes & 22676 & NaN & NaN & 2943 & 26487 & 6.70e-02 & 1.59e+04 & 27 & 22676 & \textcolor{red}{2.80e-01} & \textcolor{red}{1.52e+03} \\ 
  \hline
  \end{tabular}
  }
  \caption{Summary of simulation results on the maximin measure (the larger the better) and the MaxPro measure (the smaller the better) of the resutled designs from the candidate set generating by one-step acceptance/rejection LHDs, adaptive SCMC, and CoMinED on all benchmark problems. The values for LHDs and adaptive SCMC are average over 50 runs. CVN stands for constraint value normalization. NaN indicates no result, meaning that none of the runs generate enough candidate samples for constructing the required designs.}
  \label{tab:design_measure}
\end{table}

\end{document}